\documentclass[twocolumn]{aastex631}
\usepackage{comment}
\usepackage{amsmath, amsthm, amssymb, amsfonts, tensor}
\usepackage{MnSymbol,bbding,pifont}
\usepackage{multirow}

\newcommand\aastex{AAS\TeX}
\newcommand\latex{La\TeX}

\emergencystretch=\maxdimen
\begin{document}
\title{The \textit{JWST} Early Release Science Program for Direct Observations of Exoplanetary Systems III: \\ Aperture Masking Interferometric Observations of the Star HIP\,65426 at $\boldsymbol{3.8\,\rm{\mu m}}$}

\author[0000-0003-2259-3911]{Shrishmoy Ray} 
\affiliation{School of Mathematics and Physics, University of Queensland, St Lucia, QLD 4072, Australia}
\affiliation{University of Exeter, Astrophysics Group, Physics Building, Stocker Road, Exeter, EX4 4QL, UK}

\author[0000-0001-6871-6775]{Steph Sallum}
\affiliation{Department of Physics and Astronomy, University of California, Irvine, 4129 Frederick Reines Hall, Irvine, CA, USA}


\author[0000-0001-8074-2562]{Sasha Hinkley}
\affiliation{University of Exeter, Astrophysics Group, Physics Building, Stocker Road, Exeter, EX4 4QL, UK}

\author{Anand Sivaramkrishnan}
\affiliation{Space Telescope Science Institute, 3700 San Martin Drive, Baltimore, MD 21218, USA}

\author{Rachel Cooper}
\affiliation{Space Telescope Science Institute, 3700 San Martin Drive, Baltimore, MD 21218, USA}

\author[0000-0003-2769-0438]{Jens Kammerer}\affiliation{Space Telescope Science Institute, 3700 San Martin Drive, Baltimore, MD 21218, USA}

\author[0000-0002-7162-8036]{Alexandra Z. Greebaum}\affiliation{IPAC, Mail Code 100-22, Caltech, 1200 E. California Blvd., Pasadena, CA 91125, USA}

\author{Deeparshi Thatte}
\affiliation{Space Telescope Science Institute, 3700 San Martin Drive, Baltimore, MD 21218, USA}

\author[0000-0002-5823-3072]{Tomas Stolker}
\affiliation{Leiden Observatory, Leiden University, Niels Bohrweg 2, 2333 CA Leiden, The Netherlands}


\author[0000-0001-7819-9003]{Cecilia Lazzoni}\affiliation{University of Exeter, Astrophysics Group, Physics Building, Stocker Road, Exeter, EX4 4QL, UK}

% ...

\begin{abstract}

We present aperture masking interferometry (AMI) observations of the star HIP 65426 at $3.8\,\rm{\mu m}$ as a part of the \textit{JWST} Direct Imaging Early Release Science (ERS) program obtained using the Near Infrared Imager and Slitless Spectrograph (NIRISS) instrument. This mode provides access to very small inner working angles (even separations slightly below the Michelson limit of ${}0.5\lambda/D$ for an interferometer), which are inaccessible with the classical inner working angles of the \textit{JWST} coronagraphs. When combined with \textit{JWST}'s unprecedented infrared sensitivity, this mode has the potential to probe a new portion of parameter space across a wide array of astronomical observations. Using this mode, we are able to achieve a {5$\sigma$} contrast of ${\Delta m_{F380M}{\sim}7.62{\pm}0.13}$\,mag relative to the host star at separations ${\gtrsim}0.07\arcsec$, and the contrast deteriorates steeply at separations ${\lesssim}0.07\arcsec$. However, we detect no additional companions interior to the known companion HIP\,65426\,b (at separation ${{\sim}0.82\arcsec}$ or, $87^{+108}_{-31}\,\rm{au}$). Our observations thus rule out companions more massive than $10{-}12\,\rm{M\textsubscript{Jup}}$ at separations ${\sim}10{-}20\,\rm{au}$ from HIP\,65426, a region out of reach of ground or space-based coronagraphic imaging. These observations confirm that the AMI mode on \textit{JWST} is sensitive to planetary mass companions at close-in separations (${\gtrsim}0.07\arcsec$), even for thousands of more distant stars at $\sim$100\,pc, in addition to the stars in the nearby young moving groups and associations as stated in previous works. This result will allow the planning and successful execution of future observations to probe the inner regions of nearby stellar systems, opening an essentially unexplored parameter space.

\end{abstract}




\section{Introduction} 
\label{sec:intro}


\textit{JWST} \citep{2006Gardner, 2023Gardner} has established itself as the world's pre-eminent infrared observatory. The Near-Infrared Imager and Slitless Spectrograph instrument \citep[NIRISS,][]{DoyonNIRISS,2023Doyon}, is equipped with a sparse aperture mask \citep[][]{SivaramkrishnanNRM,2023Sivaramkrishnan}. This mask enables the Aperture Masking Interferometry \citep[AMI,][]{1986Baldwin,1987Haniff,1988Readhead} mode on the telescope, which is available for the first time from space. This facilitates the execution of science cases that can use moderate contrast at small angular separations, for example, high-resolution imaging of extended sources, structure identification of Solar System objects (e.g. Jupiter's moon Io) and imaging extremely bright objects like Wolf-Rayet stars \citep{2023Lau}. 

A particularly powerful feature of this mode is its ability to directly image planetary mass companions (PMCs) inside the Rayleigh diffraction limit, ${\lesssim}\lambda/D$ (where $D$ is the aperture size of the telescope and $\lambda$ is the observing wavelength). With the \textit{JWST}/NIRISS/AMI mode, contrasts of ${\sim}{10^{-3}}{-}{10^{-4}}$ are predicted to be achievable with sufficient integration times \citep[][]{2020SoulainSPIE,2022JWSTCommissioning}. This mode is well equipped to directly detect or place constraints on additional inner companions of systems with already-known wide-separation companions. Such systems have historically revealed the presence of additional companions at smaller orbital separations \citep[][]{2010Marois,2020Nowak,2023Hinkley}, which has also been demonstrated by theoretical studies \citep{Wagner:2019ApJ}. Using AMI to expand the parameter space inwards will hence enable the community to provide a more robust characterization of the orbital architectures of planetary systems. In addition to this, the mode is also predicted to access Jupiter mass companions at water frost-line separations around stars in nearby young moving groups and associations \citep[][]{2019Sallum,2022Ray}. Going forward, this will be a promising technique for placing constraints on initial entropies (by using in conjunction planetary mass estimates, from \textit{Gaia} or long term orbital monitoring) of planets for at least the next decade \citep[][]{2012Spiegel,2022Ray}.

The Director’s Discretionary Early Release Science (DD-ERS) Program 1386 \citep[][]{2022Hinkley} includes AMI observations to demonstrate the observing mode and test its feasibility for these science cases. In this paper we report the results of the AMI observations of the ERS program, of the star HIP\,65426 (A2V, $109.20 \pm 0.74\,\rm{pc}$), observed at $3.8\,\rm{\mu m}$. This star has a previously known companion HIP\,65426\,b \citep{2017Chauvin, 2019Cheetham}, with a projected separation of $86^{+116}_{-31}\,\rm{au}$ and a mass of $7.1{\pm}1.2\,\rm{M\textsubscript{Jup}}$ \citep{2023Carter}. 
In this paper we present the scientific results of the ERS 1386 AMI observations, placing new constraints on the presence of additional companions in the HIP\,65426 system and demonstrating its potential for similar observations in other nearby systems. 
In an accompanying paper \citep{sallum_subm} we analyze the performance, achievable contrast, and limiting noise sources of this mode. 

This work was carried out in parallel with other science aspects of the ERS program. These included (1) coronagraphic observations of HIP\,65426\,b at $2{-}5\,\rm{\mu m}$ with the Near-Infrared Camera (NIRCam) and $11{-}16\,\rm{\mu m}$ with the Mid-Infrared Instrument (MIRI) \citep{2023Carter}, (2) the highest fidelity spectrum to date of a planetary-mass object, VHS\,1256\,b \citep{2023Miles} and, (3) coronagraphy out to 15.5\,$\mu$m of HD\,141569A, a young circumstellar disk, with particular focus on sampling the disk brightness on and off the $3.0\,\mu m$ H\textsubscript{2}O ice feature (Millar-Blanchaer et al. in prep, Choquet et al. in prep).

In \S\ref{sec:JWST/AMI} we give a brief summary of the AMI mode with \textit{JWST}, followed by our observation strategy in \S\ref{sec:Observation}. In \S\ref{sec:DataReduction} we detail the various data processing steps. In \S\ref{sec:Discussion} we summarise our discussion and  in \S\ref{sec:Conclusions} we present the principal conclusions of this study.

\section{\textit{JWST} Aperture Masking Interferometry}
\label{sec:JWST/AMI}
AMI transforms a conventional telescope into an interferometric array via a pupil-plane mask \citep[][]{2001Tuthill}. This is typically accomplished using a piece of metal with holes (sub-apertures) cut out of it. The images recorded by the detector are then the interference fringes produced from the light after passing through the sub-apertures, which can be analysed using Fourier techniques.
When each baseline in an aperture mask has a unique position angle and separation, a linear relationship between pupil-plane phase differences and measured fringe phase exists. Due to this non-redundancy, Fourier observables (e.g. closure phases, squared visibilities, see \S\ref{ssec:FourierObservables}) can be calculated which are robust to first-order residual phase errors \citep[following the description in][]{2013Ireland}.

\begin{figure}
    \centering
    \includegraphics[scale=0.3]{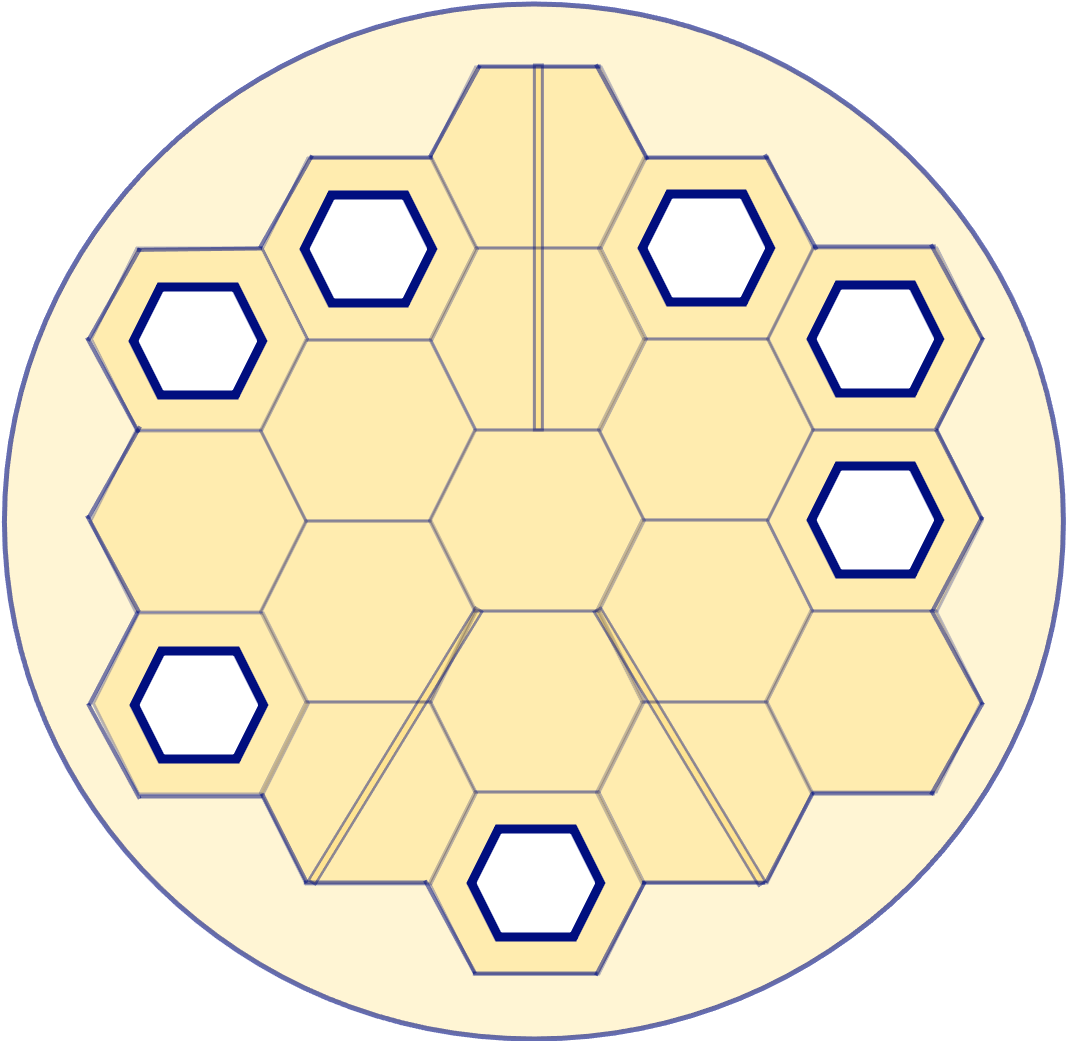}
    \caption{Schematic diagram of the non-redundant mask onboard \textit{JWST}/NIRISS containing seven sub-apertures.}
    \label{fig:NRM}
\end{figure}

This technique has been successfully carried out using ground based facilities \citep[e.g.][]{2000Tuthill,2007Monnier,2012Kraus,2015Hinkley} and is for the first time being executed in space with \textit{JWST} \citep[][]{2023Sivaramakrishnan,2023Kammerer}. This is achieved using a mask with seven hexagonal sub-apertures (see Figure \ref{fig:NRM}), each of which has an incircle diameter of $0.8\,\rm{m}$, when projected onto the \textit{JWST} primary mirror \citep{2012Sivaramkrishnan,greebaum2015}.  This probes $\left(^7_2\right)$, or 21 distinct  spatial frequencies (number of baselines) and $\left(^7_3\right)$, or 35  closure phase triangles\footnote{For a set with $n$ elements, the number of combinations with k groups is, $\left(^n_k\right)=\frac{n!}{k!(n-k)!}$. Since the mask has 7 sub-apertures, $n=7$, and $k=2$ for baselines (2 endpoints of a line segment) and $k=3$ for closure phases (3 vertices of a triangle).}. This mode can be used with four NIRISS filters at wavelengths $2.77\,\rm{\mu m}$ (F277W), $3.80\,\rm{\mu m}$ (F380M), $4.30\,\rm{\mu m}$ (F430M), and $4.80\,\rm{\mu m}$ (F480M). These wavelength channels are specifically designed to be sensitive to H\textsubscript{2}O, CH\textsubscript{4}, CO\textsubscript{2} and CO features respectively \citep[e.g. Figure 1 of][]{2020SoulainSPIE}.

\section{Observing Strategy}
\label{sec:Observation}

\begin{table*}
\centering
\caption{Parameters for the \textit{NIRISS/AMI} observations in the F380M filter. The observations were executed on 30 July 2022 (UTC). The start and the end times indicate the start and the end of the observation block at UTC and are not directly related to the exposure times. The target star, HIP\,65426 was observed between the calibrator stars  HD\,115842 and  HD\,116084.}\label{table:OBs}
\begin{tabular}{cccccccccc}
\hline
\hline

\multicolumn{1}{c}{\textbf{Star}} &  \textbf{Type} &\textbf{Start time} &\textbf{End time} &\multicolumn{1}{c}{\textbf{Sequence}} & \textbf{Readout} & \textbf{N\textsubscript{groups}}  & \textbf{N\textsubscript{ints}} & \textbf{Dithers}&\textbf{t\textsubscript{exp} (s)}     \\

\hline
\textbf{ HD\,115842  }                                      & Calibrator & 05:26:46 & 06:52:10 &    1a  & NISRAPID & 2 & 10000 & 1 & 2468.00    \\
                          &  &   &  & 1b & NISRAPID & 2 & 5500 & 1  & 1357.40  \\
\textbf{HIP\,65426}                                     & Target & 07:02:39 & 10:53:00 &        2a & NISRAPID         & 13 & 10000 & 1  &10766.40   \\
                                  &  &   &  &      2b  & NISRAPID         & 13 & 950 & 1  &1022.81  \\
\textbf{ HD\,116084}                                          & Calibrator  & 11:02:17 & 12:49:58 &  3a & NISRAPID & 3 & 10000 & 1  &3222.4  \\
                                      &  &    &   & 3b  & NISRAPID & 3 & 6000 & 1 & 1933.44    \\
\hline
\end{tabular}

\end{table*}

The AMI observations were performed using the F380M filter and the SUB80 subarray with the NISRAPID readout pattern and no dithering. The F380M filter was specifically chosen from the available filters to reach smaller angular separations than the F430M and F480M filters, to detect close-in companions whilst making sure the central wavelength was close to the region of the spectrum where exoplanets are the brightest \citep[at ${\sim}4{-}5\,\rm{\mu m}$, e.g. Figure 2 of][as compared to the F277W filter which typically spans wavelengths where directly imaged exoplanets are relatively faint]{2022Ray}.

The observing setup was designed such that a photon noise limited observation would be able to access a contrast of $\Delta_{mag}{\sim }10$ \citep{2013Ireland}. The appropriate
number of groups and integrations were hence chosen to collect ${\sim}10^{10}$ photons for each target while avoiding saturation or non-linearity effects, as presented in Table \ref{table:OBs} (see Appendix \ref{append:GroupsAndIntegrations} for an explanation of the terms `groups' and `integrations'). Table \ref{table:OBs} also presents the UTC start and end times of the observations of the target and calibrator stars. And, $\boldsymbol{t_{exp}}$ is the total exposure time for each observation sequence. All the observations were carried out following the recommended best practices as known prior to launch and described in the \textit{JWST} user documentation.\footnote{\href{https://jwst-docs.stsci.edu/jwst-near-infrared-imager-and-slitless-spectrograph/niriss-observing-strategies/niriss-ami-recommended-strategies}{https://jwst-docs.stsci.edu/jwst-near-infrared-imager-and-slitless-spectrograph/niriss-observing-strategies/niriss-ami-recommended-strategies}\label{fn-SAMpy}} 

To ensure optimal data analysis of the AMI observations, point spread function (PSF) reference stars are required to be observed, in addition to the science target. This is executed because while Fourier observables are robust to phase errors to first order, higher order errors remain that must be calibrated out \citep{2013Ireland}. These reference stars should be point sources so that they measure only the instrumental contributions to the errors in the observables, which in turn, ensures the optimal calibration of the target star observations. The target star was planned to be observed with the most time and the calibrators were chosen to be relatively bright stars, close to the target star on the sky plane. This ensured that we obtained similar number of photons from the calibrators as the target star, with shorter exposure times (see Table \ref{table:OBs}).  Each calibrator had a total exposure time of ${{\sim}1\,\rm{hour}}$ and the target star had an exposure time of ${{\sim}3\,\rm{hours}}$. However, for future observations with this mode, calibrators with similar brightness as the target star should be chosen, as the detector systematics are the limiting factor in terms of reaching the required contrast for datasets of this depth \citep[see \S \ref{ssec:FourierObservables} and][for more details]{sallum_subm}. 

For the AMI sequence of this program, the observation of the target star HIP\,65426 was preceded by one calibrator (HD\,115842) and followed by another calibrator (HD\,116084, see Section \ref{ssec:vetting} for details). Observations of two calibrator stars reduces the risk of encountering calibrators with unexpected resolved structure (e.g. close-in companions, disks). Observing the target star in the middle of the sequence between two calibrator stars which are close on the sky plane, (1) is advantageous towards reducing the wavefront drift and, (2) ensures that the time elapsed during the slew between the target and the reference stars is kept to a minimum, which minimises thermal drifts by pointing changes. This ensures that the changing spacecraft altitude does not cause a change in the temperature of the primary mirror segments and subsequently affect the Fourier phases observed by the telescope. 
The PSF reference stars are then used to calibrate out instrumental contributions to the interferometric observables (see \S \ref{sec:DataReduction}). A raw image taken with the \textit{JWST}/NIRISS/AMI mode of the science target (HIP\,65426) is shown in the left panel of Figure \ref{fig:ImagePS}. 

\subsection{Vetting of the calibrator stars}
\label{ssec:vetting}
To ensure the calibrator stars,  HD\,115842 and  HD\,116084, were point sources, they were vetted first using \texttt{Search\,Cal} \citep{2006Bonneau}, and were then followed up with observations using ground based facilities. This was firstly done with the speckle imaging observations from Southern Astrophysical Research (\textit{SOAR}) telescope's High Resolution Camera \citep[HRCam,][]{2010Tokovinin} instrument. Both calibrators (HD\,115842 and  HD\,116084) were found to be unresolved point sources. This result is presented in Table \ref{table:SOAR}. For the calibrator HD\,115842, the limiting contrasts at $0.15\arcsec$ and $1.00\arcsec$ were found to be $\Delta{m}{\sim}2.71$ and $\Delta{m}{\sim}5.00$ respectively at $0.8\,\rm{\mu m}$, and $\Delta{m}{\sim}3.66$ and $\Delta{m}{\sim}5.47$ respectively at $0.5\,\rm{\mu m}$. And for the calibrator HD\,116084, the limiting contrasts at $0.15\arcsec$ and $1.00\arcsec$ were found to be $\Delta{m}{\sim}2.97$ and $\Delta{m}{\sim}5.10$ respectively at $0.8\,\rm{\mu m}$, and $\Delta{m}{\sim}3.44$ and $\Delta{m}{\sim}5.45$ respectively at $0.5\,\rm{\mu m}$.   

\begin{table}[]
\centering
\caption{SOAR observations of the \textit{JWST} calibrators at separations $\theta_1=0.15\arcsec$ and $\theta_2=1.00\arcsec$. The quantities ${\Delta m_{\theta_1}}$ and ${\Delta m_{\theta_2}}$ are the achieved contrasts at the separations  ${\theta_1}$ and ${\theta_2}$ respectively. These stars were observed prior to the \textit{JWST} observations to check them for any potential companions or extended sources.}
\label{table:SOAR}
\hspace{-30pt}\begin{tabular}{cccc}
\hline
\hline
Calibrator & Filter  & $\Delta m_{\theta_1}$ & $\Delta m_{\theta_2}$   \\
\hline
\textbf{HD 115842}  & I ($0.8\,\rm{\mu m}$) & 2.71        & 5.00          \\
           & y($0.5\,\rm{\mu m}$)  & 3.66        & 5.47          \\
\textbf{HD 116084}  & I ($0.8\,\rm{\mu m}$) & 2.97        & 5.10          \\
           & y($0.5\,\rm{\mu m}$)  & 3.44        & 5.45          \\
\hline
\end{tabular}
\end{table}

To check the reference stars further with higher sensitivity, we also observed them with AMI observations using the \textit{VLT}/SPHERE/IFS \citep{2016Cheetham} integral field spectrograph (Proposal\,ID: 109.24EY). 
Both stars were found to be unresolved point sources, with an average contrast limit of $\Delta_{mag}{\sim}6$, calculated across the 39 IFS wavelength bins. There were no significant variations on the limit of the contrast as a function of position angle.
This contrast was achieved at separations of ${\sim}\lambda/D$ for both stars, which corresponds to separations of $23{-}40\,\rm{mas}$ given the IFS wavelength range of $0.95{-}1.60\,\rm{\mu m}$.


\section{Data Reduction and Extraction of Fourier Observables}
\label{sec:DataReduction}

\begin{figure}
    \centering
    \includegraphics[scale=0.60]{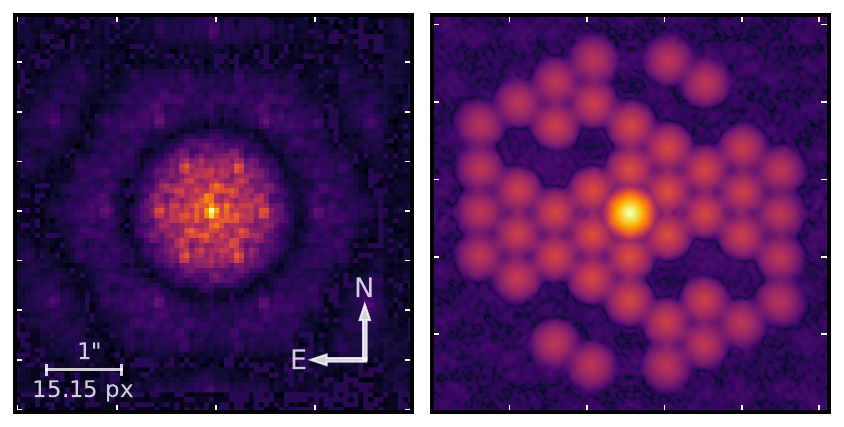}
    \caption{On the left is the science raw image of HIP\,65426. This pattern (interferogram) is obtained on the detector and is a result of the interference of the light emerging from each of the sub-apertures of the mask as shown in Figure \ref{fig:NRM}. On the right is the associated power spectrum of the science image which is the modulus squared of its Fourier transform. The power spectrum is used to extract the orbital properties of a potential companion in the star system by analysing the Fourier observables. The observations in the figure are taken at $3.80\,\rm{\mu m}$.}
    \label{fig:ImagePS}. 
\end{figure}

As part of the science enabling products produced by this ERS team, we have developed \texttt{SAMpy}\footnote{\href{https://github.com/JWST-ERS1386-AMI/SAMpy}{https://github.com/JWST-ERS1386-AMI/SAMpy}\label{fn-SAMpy}} \citep{StephSPIE}, which primarily handled the processing of this dataset, in conjunction with the \texttt{jwst}\footnote{\href{https://jwst-pipeline.readthedocs.io}{https://jwst-pipeline.readthedocs.io}\label{fn-jwst}} \citep{Bushouse2022} pipeline\footnote{All data were processed using pipeline version, \texttt{jwst}=1.7.1}. \texttt{SAMpy} is a publicly available \textsc{python} package containing data reduction tools tailored for \textit{JWST}/NIRISS/AMI, and is flexible enough to adapt to arbitrary masking setups (e.g. \textit{VLT}/SPHERE, \textit{LBT}/LMIRCam, \textit{Keck}/NIRC2). It processes the AMI data using a Fourier-plane approach. The accompanying paper \citet{sallum_subm} covers a detailed description and justification of the processing steps undertaken for this dataset, which is briefly outlined in the following sections for clarity.  

\subsection{Pre-processing of the data}
To prepare the data for the calculation of the Fourier observables, some pre-processing steps using \texttt{SAMpy} were executed. The first step in this process was to identify and correct bad pixels.
First, the bad pixel map was produced by the \texttt{jwst} stage 1 pipeline, which flags all ``DO NOT USE" pixels in the Data Quality (DQ) array as bad.
This was followed by identifying additional bad pixels using the statistics of the individual integrations and the set of integrations. 
All such bad pixels were corrected, using the Fourier-plane approach taken in \citet{2019MNRAS.486..639K}. 
Finally, each image was centred to pixel-level precision, cropping to a smaller size of $64{\times}64$ pixels, before applying a 4th-order super-Gaussian window with a FWHM of 48 pixels. This process is described in detail in \citet{sallum_subm}. 


\subsection{Data processing and model fitting}
\label{ssec:FourierObservables}
Once the processed image was obtained, the Fourier observables were calculated from the Fourier transform of the image (the power spectrum is shown in the right panel of Figure \ref{fig:ImagePS}). First, the complex visibilities were calculated which comprise the amplitudes and phases associated with the unique mask baselines. This was followed by the calculation of squared visibilities, which are the powers (amplitudes squared) associated with each of the unique mask baselines. Next, the closure phases were computed, which are the sums of phases for baselines forming a triangle. 

After the calculation of the above, the observations were calibrated. Calibrating the observations for the science target (HIP\,65426) was explored by calibrating it separately with each of the calibrator stars  HD\,115842 and  HD\,116084 respectively, in addition to calibrating it with both the calibrator stars together. It was found that using  HD\,116084 solely to calibrate the science target yielded the best results in terms of reaching the deepest contrast ($\Delta m_{F380M}{\sim}7.8$, see third column of the top panel of Figure \ref{fig:ChiSqMap}). The two likely reasons for this are that (1) this observation had more similar charge migration properties to HIP 65426 (see Section \ref{ssec:discrepancy}), and (2) used a larger number of groups (three, as opposed to two for HD\,115852). Both of these characteristics result in better minimization of PSF artefacts and detector systematics during data reduction and subsequent AMI calibration, making for a better calibrated dataset \citep{sallum_subm}. Hence, only  HD\,116084 was used for the calibration and the subsequent analysis in this paper. 

The calibration was done by (1) dividing the squared visibilities of the target by those of the calibrator and, (2) subtracting the closure phases of the calibrator from the target closure phases (see Figure \ref{fig:ChiSqMap}). 
As described in \citet{sallum_subm}, we calculate both statistical error bars and systematic error bars for the closure phases and squared visibilities.
We calculate statistical error bars by measuring the standard deviation of each quantity across all calibrated integrations.
However, these are significantly smaller than the residual calibration errors in the data. 
We thus also estimate systematic error bars by measuring the standard deviation of the closure phases and squared visibilities across the triangles and baselines, respectively. All the stars (namely, the target HIP\,65426 and the calibrators HD\,115842 and HD\,116084) were found to have no systematic variation of visibility with baseline length and hence were unresolved.

As described in more detail in the accompanying paper \cite{sallum_subm}, the closure phase scatter can have contributions from random (for example, Poisson) noise and systematic noise. As an example of connecting the standard deviations of closure phases across triangles to noise sources, we can first consider the case where Poisson noise dominates. Each closure phase would have an error bar determined by the standard error (i.e. the standard deviation of the closure phase as measured across N frames, divided by the square root of the number of frames, N). This means that the distribution of the closure phases across the different triangles would have a standard deviation equal to that same standard error (assuming that each triangle has the same standard error).

We measure the standard deviation of the closure phases across the different triangles under the assumption that the systematic errors are approximately equal for all closing triangles, and that they are (similar to the Poisson noise description above) well modelled by a Gaussian distribution. In reality, different closing triangles might have different levels of both random and systematic errors. This is because the levels of Poisson noise vary with raw visibility amplitude, and systematic errors (which have many sources) are unlikely to be uniform with spatial frequency. However, the systematic errors cannot be measured more robustly than this, given the available data.

To test whether the scatter in the calibrated data (for HIP\,65426 as well as for the two calibrators calibrated against one another) can be well modelled by a Gaussian distribution, we compared the closure phases across triangles to many realisations of random closure phases drawn from a Gaussian distribution with the same standard deviation. This procedure reproduces the data reliably. We thus determine these statistical errors to be adequate given the limitations in our ability to measure systematic errors.

This method of estimating error bars is conservative, since the scatter in the closure phases across triangles is affected by both signal and noise. We thus only adopt these error bars after determining that the best fit (corresponding to the least $\chi^2$ value) to the dataset is indeed caused by noise. We do this first by examining the quality of the best fit by calculating its reduced $\chi^2$ value if only the standard errors were used, which for the fit to the HIP\,65426 data calibrated with HD\,116084 was calculated to be ${\sim}41$. This high reduced $\chi^2$ value implies that the errors are indeed underestimated. Hence, larger errors on closure phases ({${\sim}0.14^{\circ}$}) were adopted by calculating the standard deviation, as previously mentioned, across all the measured closure phases for this calibration.  This produced a reduced $\chi^2$ of  ${\sim}1$ for the best fit (see Figure \ref{fig:ChiSqMap}). These adopted errors were used for all our analyses in this work.

\subsection{Companion Model Fitting and Injection Tests}
\label{sec:CompanionModelFitting}

\begin{figure*}
    \centering
    \includegraphics[scale=0.9]{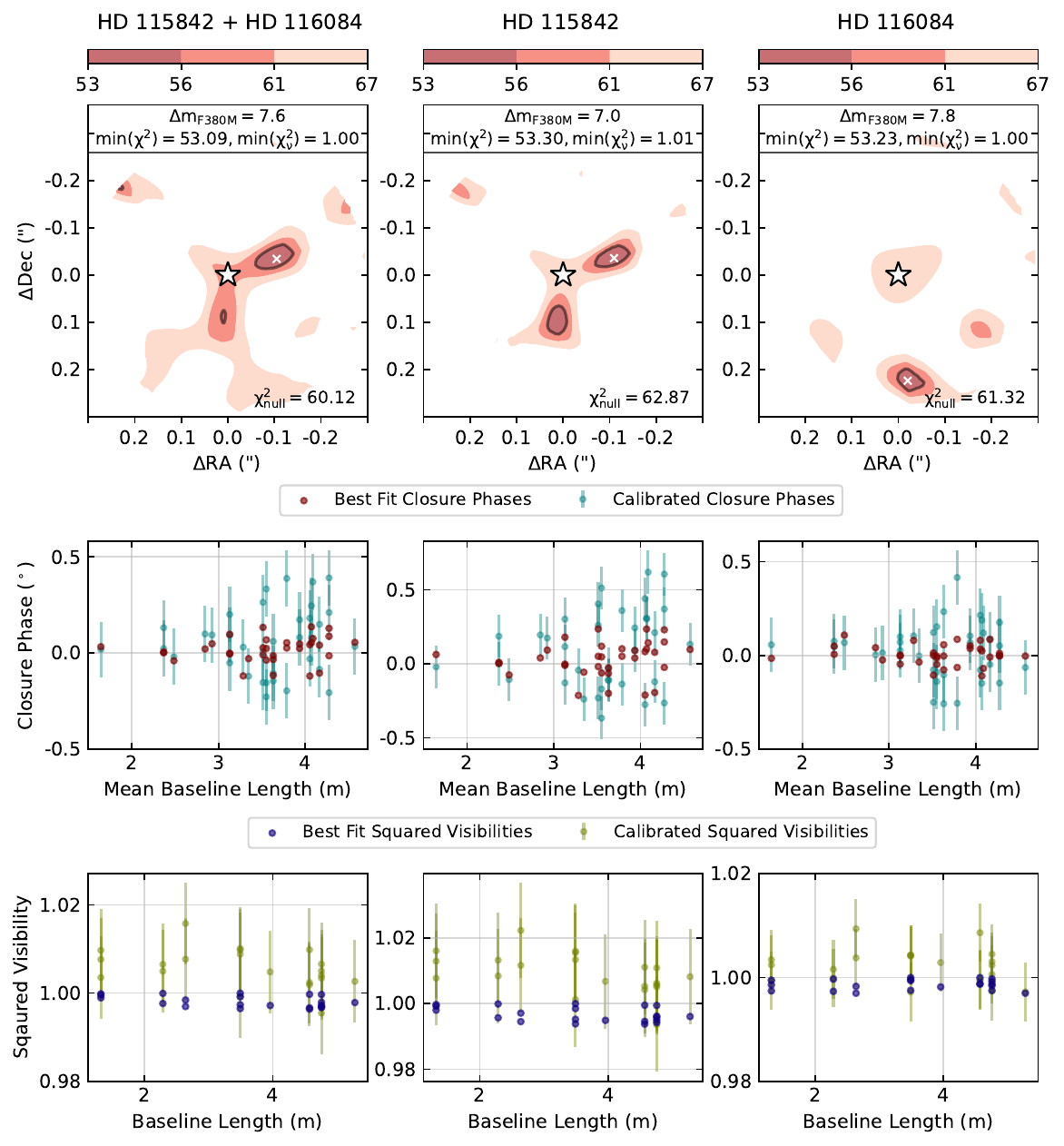}
    \caption{The first column shows the HIP\,65426 data calibrated with both calibrators (HD\,115842 and HD\,116084), the second column shows the data calibrated with only HD\,115842, and the third column shows the data calibrated with only HD\,116084. The first row shows the $\chi^2$ surfaces which are slices of the best fit contrast containing the lowest $\chi^2$ from a grid search (with parameters of companion separation, position angle and contrast) in right ascension and declination. This is shown for each HIP\,65426 calibration. The second and third rows show the corresponding best fit Fourier observables of closure phases and squared visibilities respectively, compared to the data. The `{\Large $\filledstar$}' symbols in the $\chi^2$ maps in the first row show the position of the central star and the `$\times$' symbols show the best fit companion positions for each calibration. The best fit $\Delta{m_{F380M}}$ contrast value for each calibration is also displayed in the panels of the first row. Along with this, the minimum $\chi^2$ values and the minimum reduced $\chi^2$ values (denoted by $\chi^2_{\nu}$) are also displayed. Contour regions show the $1\sigma$, $2\sigma$ and $3\sigma$ regions respectively from the lowest $\chi^2$ value. The $1\sigma$ regions are demarcated with a black contour for clarity. As discussed in Section \ref{sec:CompanionModelFitting}, we conclude that no companions are detected based on: (1) the inconsistent companion solutions from calibration to calibration, (2) the lack of a single clearly-defined region of low $\chi^2$ value for individual calibrations, and (3) the similarity of these fit result values to those for simulated noise (see Figure \ref{fig:InjectionsAndNoise}).}
    \label{fig:ChiSqMap}
\end{figure*}

\begin{table*}
\caption{This table shows the best fit (equivalently the values corresponding to the least $\chi^2$ positions in the $\chi^2$ maps) parameters from the grid search of the reduced data for different calibrations. The corresponding errors for each value is the $1\sigma$ error from the best fit as shown in Figure \ref{fig:ChiSqMap}.}
\hspace{50pt}\label{table:1sigmaerrors}\begin{tabular}{cccc}
\hline
\hline
                     & \multicolumn{1}{l}{\multirow{2}{*}{\textbf{\begin{tabular}[c]{@{}l@{}}HD~115842 +\\ HD~116084\end{tabular}}}} &                            \\
\textbf{}            &                                                                  & \textbf{HD~115842}          & \textbf{HD~116084}          \\
\hline
\textbf {Position Angle ($^\circ$)} & \hspace{18pt} $288.00^{+10.29}_{-10.28}$                                                                                  & $288.00^{+10.29}_{-2.02}$ & $185.14^{+5.88}_{-1.54}$ \\
            \textbf{Contrast ($\Delta m_{F430M}$)} & \hspace{18pt}$7.6^{+1.4}_{-0.8}$                                                                                  & $7.0^{+1.0}_{-0.6}$ & $7.8^{+1.0}_{-0.2}$ \\
  \textbf{Separation ($\arcsec$)}                                               & \hspace{18pt}$0.11^{+0.04}_{-0.49}$                                                                                  & $0.12^{+0.03}_{-0.06}$ & $0.23^{+0.02}_{-0.03}$\\
\hline

\end{tabular}
\end{table*}

\begin{figure*}
    \centering
    \includegraphics[scale=0.9]{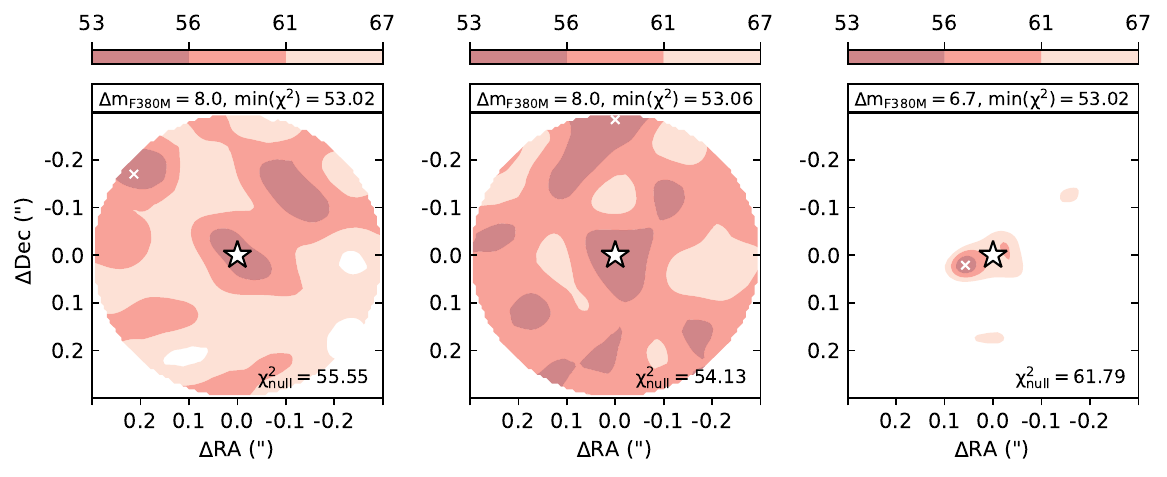}
    \caption{These plots show the $\chi^2$ maps for three Gaussian distribution simulations, with the same mean and standard deviation as that measured for the dataset HIP\,65426 calibrated  with HD\,116084. Contour regions show the $1\sigma$, $2\sigma$ and $3\sigma$ regions respectively from the lowest $\chi^2$ value similar to Figure \ref{fig:ChiSqMap}. At the top of these plots, is the best fit $\Delta m_{F380M}$ contrast value along with the minimum value of $\chi^2$ in the map. This minimum $\chi^2$ location in the above plots is denoted by `$\times$'. These contrast values approximately overlap with the best-fit contrasts for the HIP\,65426 companion fit, and the $\chi^2$ surfaces display the same lack of a single, clearly-defined low minima. Together these factors demonstrate that the fit results for HIP\,65426 can be caused by noise alone. This further demonstrates that the statistically insignificant companion signal detected in Figure \ref{fig:ChiSqMap} are likely noise artefacts in the data rather than indicating the existence of an actual companion.}
    \label{fig:InjectionsAndNoise}
\end{figure*}

\begin{figure*}
    \centering
    \includegraphics[scale=0.8]{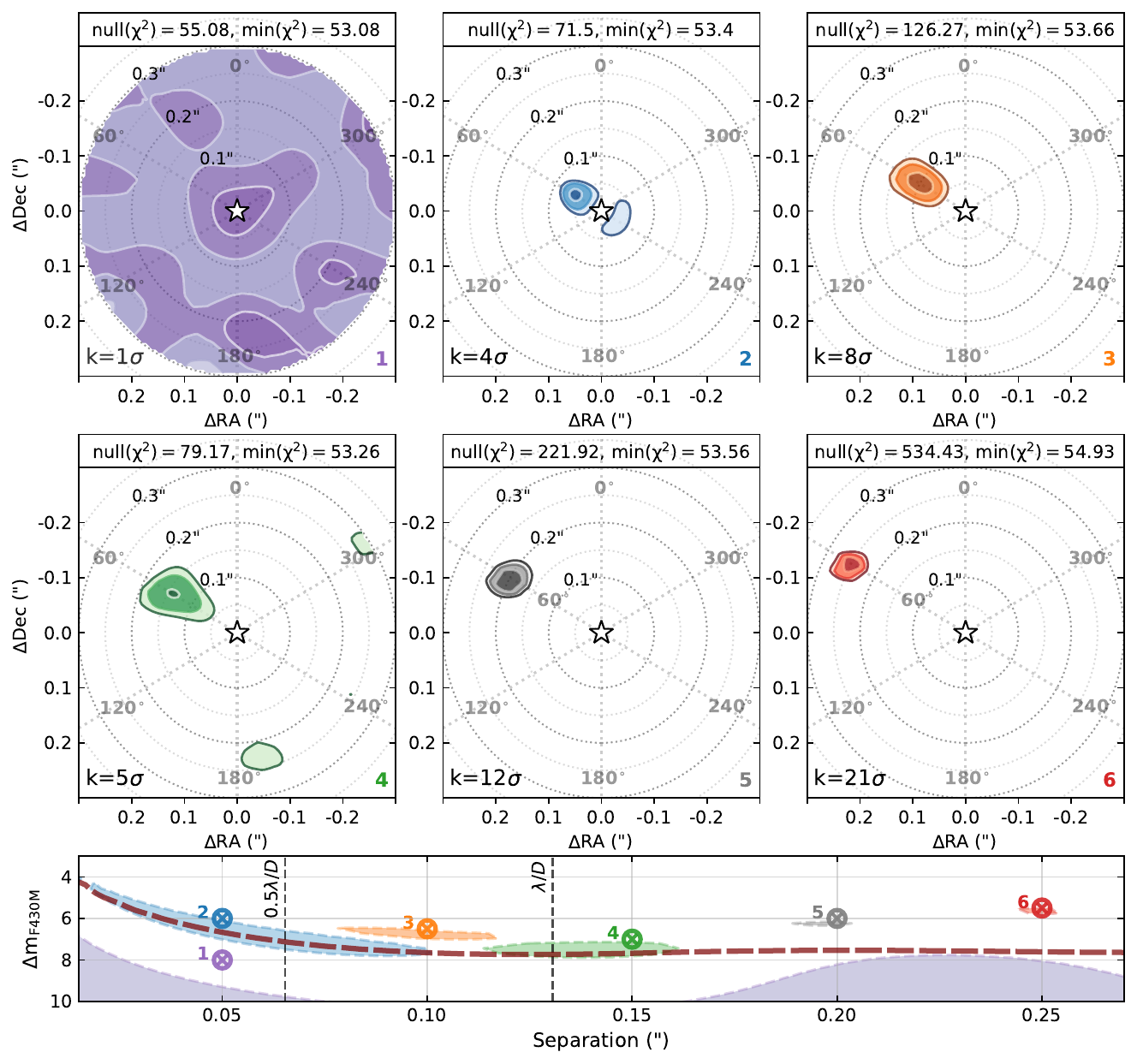}
    \caption{The plots in the first two rows show the $\chi^2$ maps at the best fit contrast surface slice with companions injected into the calibrated HIP\,65426 data at a fixed position angle ($\theta=60^{\circ}$, where $\theta$ is calculated counter-clockwise with respect to the vertical axis) with varying separations and contrasts for six {tests}. The null {(no companion)} $\chi^2$ value and the minimum $\chi^2$ values are provided at the top of each of these plots. The {test} number is provided at the bottom right of these plots. The location of these injections (in the separation vs contrast space) is shown in the bottom plot with number and colour coded `$\otimes$' symbols. The detection significance ($k$) of the best fit compared to the null model in each of the plots in the first two rows is given at the bottom left. The grid of concentric circles in the top two panels denote the separations from the host star (null $\chi^2$ model). For cases with $k{>}3\sigma$ ({tests} 2, 3, 4, 5, and 6), the contour regions are $k\sigma$, $(k-1)\sigma$ and $(k-2)\sigma$ from the null model value respectively. For {test} 1 with $k{=}1\sigma$, the contours are the $1\sigma$, $2\sigma$ and $3\sigma$ from the minimum $\chi^2$ value. The bottom panel shows the $1\sigma$ region from the best fit separation and contrast point for each of the injection {tests} at the best fit position angle slice. The X-axis showing the separation is the equivalent of the concentric circles' separation grid in the above plots. The {brown} dashed line shows the calculated $5\sigma$ contrast curve from the HIP\,65426 calibrated data (see Figure \ref{fig:ContrastCurve}). It is evident that injections near the contrast limit, companions can be recovered with a ${\sim}5\sigma$ confidence (even below separations of $0.5\lambda/D$ for {test} 2, however this causes a larger uncertainty in retrieved contrast and separation as seen in the bottom panel). At contrasts brighter than this, the recovered signal significance is considerably larger. However, at separations below $0.5\lambda/D$ and contrasts dimmer than the $5\sigma$ limit, the companion cannot be found ({test} 1 with $k{=}1\sigma$ showing a separation/contrast degeneracy in the bottom plot). These results are summarised in Table \ref{table:injections}.}
    \label{fig:ContrastsChisqComp}
\end{figure*}

\begin{center}
\hspace{-100pt}\begin{table*}[]
\caption{Summary of the injection and recovery tests shown in Figure \ref{fig:ContrastsChisqComp} with their test numbers (increasing in injected separation from test 1 to test 6). The retrieved values are the best fit values from the grid search. All the retrieved values are listed with ${1\sigma}$ errors from the best fit, rounded off to two decimal places. $k$-value is the integer detection significance at the least $\chi^2$ value ($\chi^2_{min}$) compared to the null (no companion) $\chi^2$ value ($\chi^2_{null}$). 
The injections for all the tests were made at position angle $\theta{=}60.00^\circ$. The retrieved position angle for all tests was $\theta{=}58.78^\circ$, with the exception of test 1, which had a retrieved position angle of $\theta{=}191.02^\circ$. 
}
\label{table:injections}
\hspace{-20pt}\begin{tabular}{clccccccll}
\hline
\hline 
\multicolumn{2}{c}{}  & \multicolumn{2}{c}{\textbf{\underline{Injected companion}}}                                  & \multicolumn{2}{c}{\textbf{\underline{Retrieved companion}}}                                 & \multicolumn{2}{c}{}            &                       &                               \\
\multicolumn{2}{l}{}  & \multicolumn{1}{l}{\textbf{$\boldsymbol{\Delta m_{F380M}}$}} & \multicolumn{1}{l}{\textbf{Separation($\arcsec$)}} & \multicolumn{1}{l}{\textbf{$\boldsymbol{ \Delta m_{F380M}}$}} & \multicolumn{1}{l}{\textbf{Separation($\arcsec$)}} & $\boldsymbol \chi^2_{null}$ & \multicolumn{1}{l}{$\boldsymbol \chi^2_{min}$} & {$\boldsymbol \Delta \chi^2$}&$\boldsymbol k$-\textbf{value}                       \\
\hline
\multicolumn{2}{c}{\textbf{1}} & 8.00                          & 0.05                            & $8.64^{+3.36}_{-1.69}$                          & $0.23^{+2.77}_{-0.23}$                             & 55.08   & 53.08                       & 2.00 &\multicolumn{1}{c}{$1\sigma$} \\ 
\multicolumn{2}{c}{\textbf{2}} & 6.00                          & 0.05                            & ${6.72^{+0.66}_{-2.41}}$                         & ${0.06^{+0.03}_{-0.04}}$                            & 71.50   & 53.40                       & 17.10 & \multicolumn{1}{c}{$4\sigma$} \\ 
\multicolumn{2}{c}{\textbf{3}} & 6.50                          & 0.10                            & ${6.72^{+0.02}_{-0.52}}$                          & ${0.10^{+0.02}_{-0.02}}$                           & 126.27   & 53.66                       & 72.61 & \multicolumn{1}{c}{$8\sigma$}  \\ 
\multicolumn{2}{c}{\textbf{4}} & 7.00                          & 0.15                            & ${7.44^{+0.46}_{-0.47}}$                         & ${0.14^{+0.02}_{-0.03}}$                             & 79.17   & 53.26                       & 25.91 &  \multicolumn{1}{c}{$5\sigma$} \\ 
\multicolumn{2}{c}{\textbf{5}} & 6.00                          & 0.20                            & ${6.24^{+0.16}_{-0.25}}$                           & ${0.20^{+0.00}_{-0.01}}$                            & 221.92   & 53.56                       & 168.36 & \multicolumn{1}{c}{$12\sigma$} \\ 
\multicolumn{2}{c}{\textbf{6}} & 5.50                          & 0.25                            & ${5.52^{+0.27}_{-0.03}}$                          & ${0.25^{+0.00}_{-0.01}}$                          & 534.43   & 54.93                       & 479.50 & \multicolumn{1}{c}{$21\sigma$} \\ 
\hline
\end{tabular}
\end{table*}
\end{center}

\begin{figure*}
    \centering
    \includegraphics[scale=0.6]{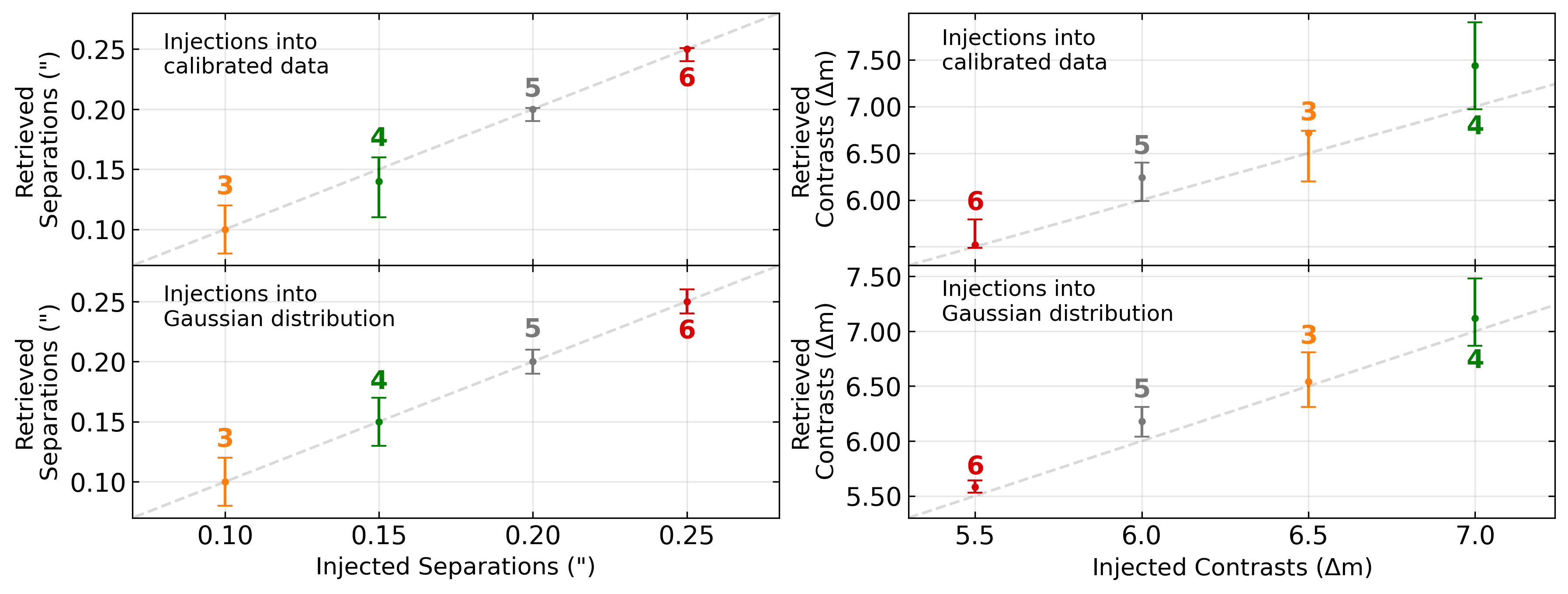}
    \caption{{ The plots show the correlation between the injected and retrieved values of companion injections. The left and the right panels show the separation and contrast values respectively. And, the top and bottom panels show the injections into calibrated data (the same as in Figure \ref{fig:ContrastsChisqComp}) and Gaussian distribution (similar to Figure \ref{fig:InjectionsAndNoise}) respectively. The injection tests are at the same contrast and separation values as in Figure \ref{fig:ContrastsChisqComp}, with their respective (color coded) test numbers for tests with $k$-values ${\geq}5\sigma$ (see Table \ref{table:injections}). The error bars on the plot points are the $1\sigma$ errors from the best fit. The retrieved values in the top panel are consistent with the injection values within $1\sigma$ (errorbars overlap with the 1:1 straight line, shown in grey) where the injections are into calibrated data. This is also true in the bottom panel where injections are carried out into Gaussian distribution of closure phases and squared visibilities, equivalent to the method used to produce Figure \ref{fig:InjectionsAndNoise} and discussed in Section \ref{sec:CompanionModelFitting}.}}
    \label{fig:GuassianDistributionInjection}
\end{figure*}

\begin{figure}
    \centering
    \includegraphics[scale=0.6]{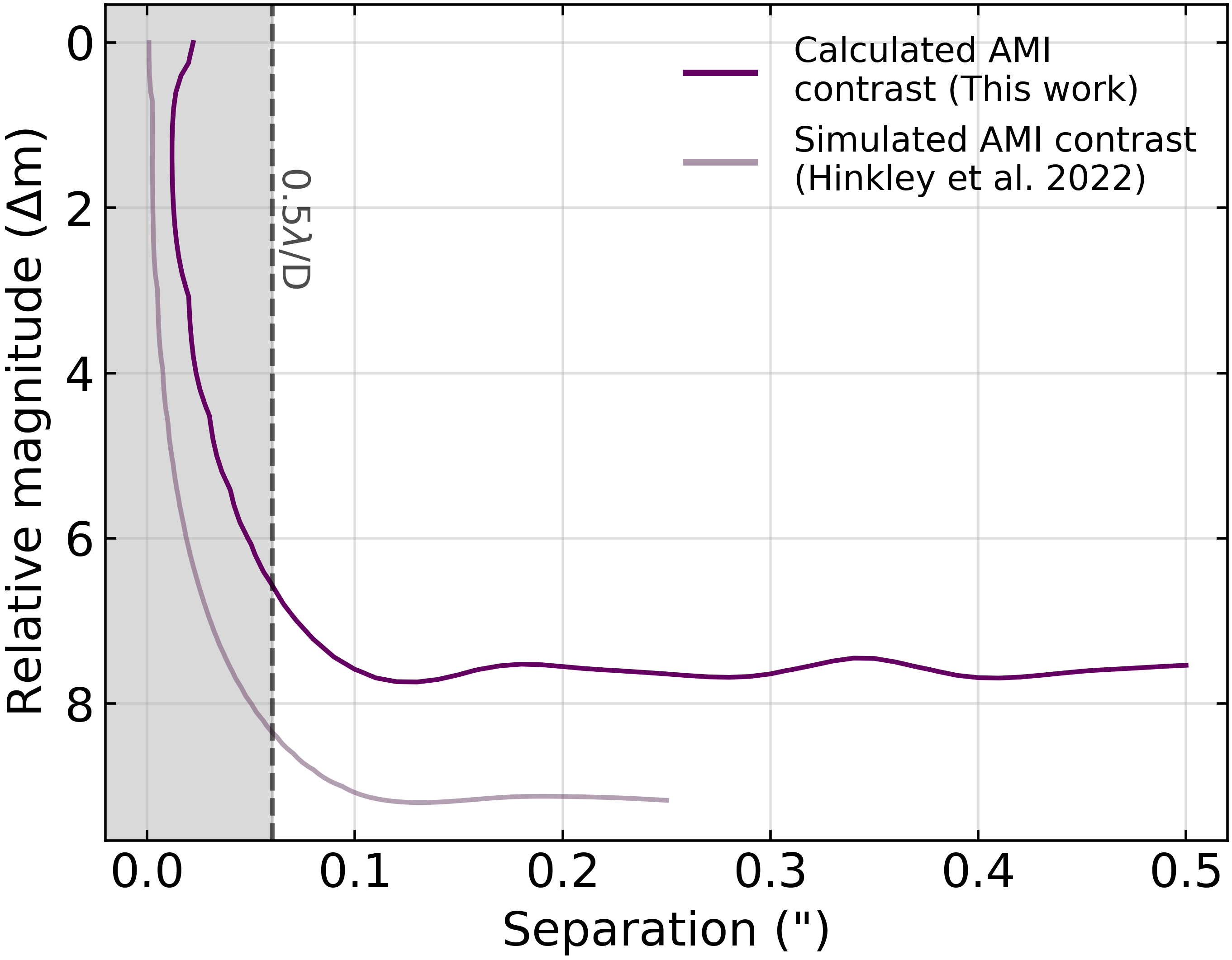}
    \caption{The calculated $5\sigma$ contrast curve for our observation is shown in purple, which was obtained using closure phases. The analogous simulated observation is shown in light purple, which was optimistic by ${\sim}1{-}2$ mag. The inner working angle for AMI observations is essentially $0.5\lambda$/D (Michelson diffraction limit). Separations below this limit are therefore greyed out. {Calculating the mean and the standard deviation of the contrast curve at separations ${\geq}0.5\lambda/D$, gives a value of $\Delta m{=}7.62{\pm}0.13$. This is the contrast we report for this work.}} 
    \label{fig:ContrastCurve}
\end{figure}

\begin{figure}
    \centering
    \includegraphics[width=\columnwidth]{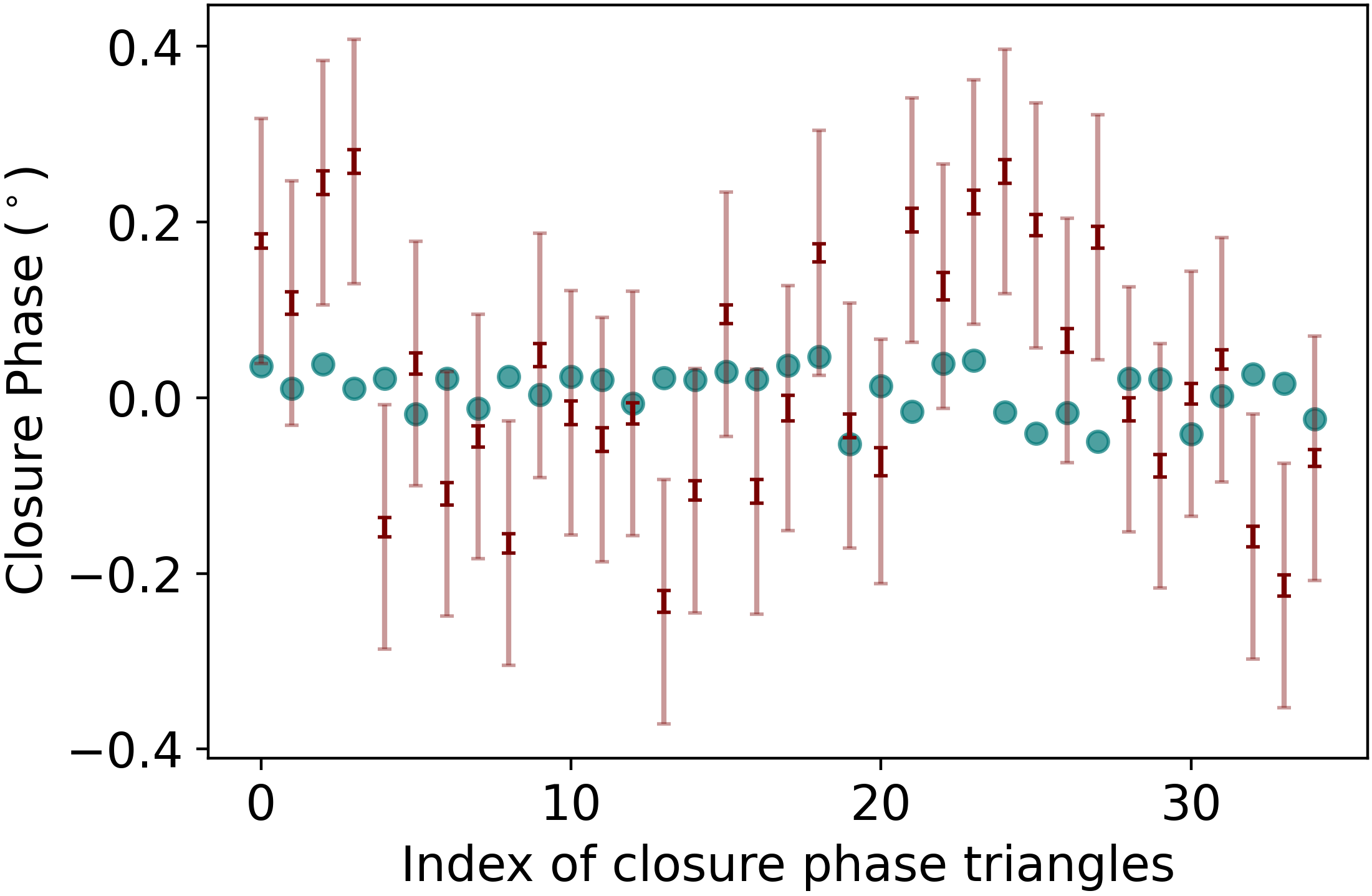}
    \caption{The expected closure phase signal from the known planet HIP\,65426\,b is shown with cyan circles and the maroon lines are the measured closure phases from the observation. The smaller error bars are the statistical errors (calculated by measuring the scatter around the mean) and the larger error bars are the adopted errors (as discussed in section \ref{ssec:FourierObservables}).}
    \label{fig:AliasedSignal}
\end{figure}

We fit companion models to the calibrated Fourier observables on the calibrated dataset via both grid and Markov-Chain Monte Carlo (MCMC) methods \citep[utilising the \texttt{emcee} \textsc{Python} package,][]{2013_Foreman-Mackey}. The analytic models consist of a delta function central point source (representing
the star) and an ensemble of delta function companions each with a separation, position angle, and
contrast relative to the central star, individually selected with the aim of obtaining a binary model (containing the central star and one companion) most closely resembling the calculated Fourier observables. This was achieved by varying the values of separation, position angle, and contrast of the companion delta function. We then take the lowest $\chi^2$ value as the best fit companion model. This $\chi^2$ value of the best fit should be significant\footnote{For the purposes of this work, we state values at ${\geq}5\sigma$ from the null value to be statistically significant.} compared to the null model for a confident detection. 

To visualise this, we plot the $\chi^2$ surface at the best fit contrast slice (the one with the lowest $\chi^2$ value) from the grid for each calibration in Figure \ref{fig:ChiSqMap}. We also plot the $1\sigma$, $2\sigma$ and $3\sigma$  contour regions from the lowest $\chi^2$ value. Since the companion model has 3 parameters, these are at values greater by 3.53, 8.02 and 14.16 respectively than the lowest $\chi^2$ value. We also state the reduced minimum $\chi^2$ values (denoted by $\chi^2_{\nu}$) in the figure\footnote{$\chi^2_{\nu}=\frac{\chi^2}{\kappa}$, where $\kappa{=}53$, the number of degrees of freedom. This is calculated as $\kappa{=}35{+}21{-}3$, where the first, second and third terms are the number of closure phases, number of squared visibilities and the number of parameters (namely position angle, separation and contrast) used in the grid search respectively.}. The typical values of the minimum $\chi^2$ and the null $\chi^2$ for the calibrated data were calculated to be ${\sim}53$ and ${\sim}61$ (see top panel of Figure \ref{fig:ChiSqMap}) respectively. Hence, the best fits in each calibration are at values ${<}3\sigma$ from the null model, which is a statistically insignificant companion signal. The results of this reduction for each calibration is summarised in Table \ref{table:1sigmaerrors}. The best fit (least $\chi^2$) values are presented with the $1\sigma$ error bars for each parameter.

While a best-fit model can be found for each dataset, the best-fit models differ from one calibration to another, with contrasts between $\Delta m_{F380M} = 7.0{-}7.8$ and widely varying separations and position angles. 
Furthermore, in all three calibrations, a unique clearly defined region of low $\chi^2$ value does not exist.
It should be noted that the method for estimating systematic error bars (measuring the standard deviation of the time-averaged closure phases and squared visibilities) causes the reduced $\chi^2$ values of the best fits in Figure \ref{fig:ChiSqMap} to be close to one. 
This does not necessarily represent the quality of the fit, but rather the large systematic uncertainties in the calibrated data.
Thus, the inconsistent solutions from calibration to calibration, the lack of a clearly-defined single $\chi^2$ minimum  and the best fit values being statistically insignificant (${<}3\sigma$) when compared to the null model in any individual calibration, argue against the existence of additional companions in this dataset. 

We used noise and companion injection simulations to explore this further, specifically for the HD,116084 calibration, since it yielded the lowest scatter and thus the deepest achievable contrast.
We first performed fits to simulated Gaussian distribution. 
  These were the closure phases and squared visibilities drawn from Gaussian distributions with standard deviations equal to those measured in the calibrated data. The mean of the closure phase distribution was set equal to zero, and the mean of the squared visibility distribution was set to be the median value of the calibrated squared visibilities.

Figure \ref{fig:InjectionsAndNoise} shows the resulting $\chi^2$ maps at the best-fit companion contrast for  three different realisations.
The best-fit contrasts, which range from $\Delta m_{F380M}{=}6.7{-}8.0$, overlap with the $\Delta m_{F380M}{=}7.8$ mag result for HIP\,65426. The typical values of minimum $\chi^2$ (best fit) in these are ${\sim}53$ and the null model values are ${\sim}54{-}61$. The best fits are hence at values ${<}3\sigma$ from the null model for all realisations, similar to the best fits of the calibrated data. This shows that the companion fit result shown in the right column of Figure \ref{fig:ChiSqMap} can be caused by noise alone. Specifically, companion contrasts of $\Delta m_{F380M}{=}8.0$ are unreliable  at wide separations (as seen in the left and centre panels of Figure \ref{fig:InjectionsAndNoise}), and $\Delta m_{F380M}{=}6.7$ at separations well within $\lambda/D$ (${\sim}$at $0.5\lambda/D$, as seen in the right panel of Figure \ref{fig:InjectionsAndNoise}). The unreliability arises from the fact that there is no statistically significant unique low $\chi^2$ region. These fits to Gaussian distributions are consistent with the calculated contrast curve in Section \ref{sec:ccurves} (also see Figures \ref{fig:ContrastsChisqComp} and \ref{fig:ContrastCurve}), since they are all below the curve given their separations.

For clarity, in Figure \ref{fig:ContrastsChisqComp}, we show $\chi^2$ maps of the calibrated (with HD\,116084) data (HIP\,65426) with injected companions. This was done by using a companion model constructed with a given value of separation, contrast and position angle. The Fourier observables (closure phases and squared visibilities) of this model were injected into the calibrated Fourier observables (closure phases and squared visibilities) of the data. {This was achieved by adding the closure phases and multiplying the squared visibilities of the model and the data.} The resulting  closure phases and squared visibilities are presented in Appendix \ref{append:closurephases} and Appendix \ref{append:squaredvisibilities} respectively. The errors for this were set to be the same as the errors of the calibrated data. Companions were injected at a position angle  of $\theta{=}60^{\circ}$ (calculated counter-clockwise with respect to the vertical axis) with varying contrasts and separations for six {tests}. Subsequently, a recovery was performed following the same approach as the real data. In Figure \ref{fig:ContrastsChisqComp}, the detection $\sigma$ significance is given by $k$ (analogous to the $\Delta\chi^2$ value discussed in later in Section \ref{sec:ccurves}, $k=int(\sqrt{\Delta\chi^2})=int(\sqrt{\chi^2_{null}-\chi^2_{min}})$ ) for each injection test. For {tests} with values of $k{>}3\sigma$, the contour regions plotted are at $k\sigma$, $(k-1)\sigma$ and $(k-2)\sigma$ from the null {(no companion)} $\chi^2$ value.  

The injected companions with $m_{F380M}{\lesssim}7.5$ {(the contrast limit from the 5$\sigma$ contrast curve discussed in Section \ref{sec:ccurves}, shown in Figure \ref{fig:ContrastCurve})} at separations beyond the Michelson diffraction limit ($0.5\lambda/D$), are clearly detected as distinct regions of low $\chi^2$ values (with $k{\geq}5\sigma$) with preferred positions {(in {tests} 3, 4, 5 and 6) in the first two rows of Figure \ref{fig:ContrastsChisqComp}}, unlike the noise realisations and the calibrated data (in Figures \ref{fig:ChiSqMap} and \ref{fig:InjectionsAndNoise}).
Decreasing the companion contrast to levels of $\Delta m_{F380M}{>}7.5$, eventually causes the resulting $\chi^2$ surfaces to become indistinguishable from noise, as in the case of {test} 1 in Figure \ref{fig:InjectionsAndNoise}. For this case, the contour regions are plotted at $1\sigma$, $2\sigma$ and $3\sigma$ intervals from the minimum $\chi^2$ value (similar to the plots in Figures \ref{fig:ChiSqMap} and \ref{fig:InjectionsAndNoise}). 
In {test} 2, a relatively bright companion ($\Delta m_{F380M}{\sim}6.0$) is injected below the Michelson diffraction limit ($0.5 \lambda/D$), and hence the recovered signal is not quite as significant ($k{=}4\sigma$) as the other cases (with ${>}0.5 \lambda/D$, see Section \ref{sec:ccurves} for further discussion). These injection and recovery tests illustrated in Figure \ref{fig:ContrastsChisqComp}, are summarised with the relevant values in Table \ref{table:injections}. {The retrieved values of contrast and separation are the best fit values obtained from a grid search. And the errors reported on these values in the table are the $1\sigma$ errors from the best fit. The corresponding regions are shown in the bottom panel of Figure \ref{fig:ContrastsChisqComp}.} 

{We also show the correlation plots of the injection and recovery tests (with $k$-values ${\geq}5\sigma$ as reported in Table \ref{table:injections}) with their $1\sigma$ errors in the top panel of Figure \ref{fig:GuassianDistributionInjection}. As is evident from the figure, all these tests (with test numbers 3, 4, 5, and 6) overlap with the 1:1 line within the $1\sigma$ errorbar, for both separation and contrast values. This establishes that the retrieved values are consistent with the injected values for companions when the signal is detected with a ${\geq}5\sigma$ significance. In the bottom panel of Figure \ref{fig:GuassianDistributionInjection}, we perform the same injections as the top panel to a Gaussian distribution. This distribution was produced in the same way as Figure \ref{fig:InjectionsAndNoise} as discussed earlier in this section. In this case, the companion separations and contrasts are also retrieved consistently within the $1\sigma$ errors (i.e. they are on the 1:1 dashed grey line, similar to the injection and recovery tests in the calibrated data in the top panel). It is to be noted that the best fit points in Figure \ref{fig:GuassianDistributionInjection} are not on the 1:1 line in any case. This is due to the fact that it is not possible to perfectly recover an injected signal with infinite precision given that the noise is always going to introduce some error. This is seen in the bottom panel of Figure \ref{fig:GuassianDistributionInjection} because the Gaussian distribution does not capture the systematic errors in the actual data as is the case for with all AMI observations \citep{2013Ireland}. And this is seen in the top panel of Figure \ref{fig:GuassianDistributionInjection} as well, with non-Gaussian noise in the data. Although the retrieved values are consistent with the injected signals, the residual systematic errors in the data biases the best fit values of contrast and separation (which is expected given the significant non-Gaussian calibration errors). For this reason, we use the null model as a contrast curve estimator. This is further discussed in Section \ref{sec:ccurves}.

Lastly, we compare the HIP\,65426 data calibrated with HD\,116084 to expectations for the signal from the known companion in the system, HIP\,65426\,b \citep{2017Chauvin,2023Carter}.
Figure \ref{fig:AliasedSignal} shows the results. 
The expected closure phase signal from HIP\,65426\,b is shown with cyan circles. 
The calculated closure phase signal from the observation is shown in maroon with error bars. The smaller error bars are the statistical errors calculated from the science and calibrator observations added in quadrature. The longer error bars are the standard deviation of the calibrated closure phases of the science target (the difference between the science and the calibrator observations, see \S \ref{ssec:FourierObservables}).
Spatial frequency of NIRISS/AMI is capable of detecting the aliased signal from the companion. However, Figure \ref{fig:AliasedSignal} shows that the signal from HIP\,65426\,b is at a significantly lower level than the scatter in the calibrated data, making the known companion undetectable.


\subsection{Accessible Companion Contrast}\label{sec:ccurves}
A contrast curve was generated from
a single-companion fit model. This was executed following an approach similar to \cite{2019Sallum}, which is briefly discussed here for clarity \citep[see also][]{sallum_subm}. The calibrated closure phases and squared visibilities were compared to the closure phases and squared visibilities of a grid of single companion models with different separations, contrasts, and position angles. For each companion separation, the average $\chi^2$ value of all the sampled position angles ($\chi^2_{sep}$) was calculated. Finally, the $5\sigma$ contrast was taken to be the contrast at which $\Delta\chi^2 = \chi^2_{null}{-}\chi^2_{sep}=25$, where $\chi^2_{null}$ is the $\chi^2$ calculated for the null (no companion, or equivalently the position of the star in the $\chi^2$ maps) model. 
The $\Delta\chi^2$ value of 25 is taken since one parameter (contrast) determines the difference between the null model and the 5$\sigma$ detectable model at each separation \citep[refer to][for a detailed description of the contrast curve calculation]{sallum_subm}. {This threshold value of $\Delta\chi^2{=}25$ is appropriate for the model selection as this corresponds to a 5$\sigma$ significance in the case of a model with 1 degree of freedom (difference in the number of parameters between the companion model and the null model).} This 5$\sigma$ curve is shown in Figure
\ref{fig:ContrastCurve} in dark purple. {Calculating the mean and the standard deviation of the contrast curve at separations ${{\geq}}0.065\arcsec$ (equivalently ${{\geq}}0.5\lambda/D$, the Michelson diffraction limit), gives a value of $\Delta m=7.62{\pm}0.13$. This is the value we quote for the achievable contrast in this work.} 

The contrast curve is also shown in Figure \ref{fig:ContrastsChisqComp} in the bottom panel with a dashed {brown} line. As mentioned briefly in Section \ref{sec:CompanionModelFitting}, all injections above this contrast curve (and separations greater than the  Michelson diffraction limit of $0.5\lambda/D$) are detected with ${\geq}5\sigma$ significance. In Figure \ref{fig:ContrastsChisqComp}, {tests} 3, 5 and 6 have high $k$ values since all these companions were injected at contrasts brighter than the $5\sigma$ contrast limit at the injected separations. {Test} 4 had a companion injected ${\sim}$at the $5\sigma$ contrast limit and hence is retrieved with a $5\sigma$ confidence at the injected separation and position angle. There are other regions of low $\chi^2$ values (for example at separations ${\sim}0.2\arcsec{-}0.3\arcsec$ with position angles ${\sim}190^\circ$ and ${\sim}300^\circ$) for this {test} in the plot (first plot in the second row of Figure \ref{fig:InjectionsAndNoise}) at ${\sim}3\sigma$ significance from the null $\chi^2$ value. This is caused by the residual systematic errors in the data.

{Test} 2 returns a companion with a relatively high confidence ($k{=}4\sigma$). However since this injection is at a separation lower than the Michelson diffraction limit, the grid search struggles to fit the contrast and separation with unique values. This manifests itself in a degeneracy which can be seen in the bottom panel of Figure \ref{fig:InjectionsAndNoise} ($1\sigma$ region from the best fit at the lowest $\chi^2$ position angle slice, shown in blue for {test} 2). This means that although we can detect the presence of a companion with a contrast greater than the $5\sigma$ contrast limit at this separation, we cannot precisely retrieve its orbital location or contrast. For this reason, the corresponding mass limits from this contrast curve were calculated with a lower limit on separation as $0.5\lambda/D$ (see Section \ref{ssec:masslimits} and Figure \ref{fig:MassLimits}).

In addition to the calculated $5\sigma$ contrast curve, Figure \ref{fig:ContrastCurve} shows the simulated contrast curve of this observation from \cite{2022Hinkley} in light purple based on pre-launch expectations of systematic noise.


\section{Discussion}
\label{sec:Discussion}
As evident from Figure \ref{fig:ContrastCurve}, the contrast curve from actual data (this work) underperforms compared to the simulated contrast curve of the observation \citep{2022Hinkley}. Below we briefly discuss possible reasons for this discrepancy, which are explored in quantitative detail in the companion paper \citet{sallum_subm}. We also discuss how the accessible contrast in magnitudes translates to the accessible mass limits based on different evolutionary models. 

\subsection{Discrepancy with Simulations}
\label{ssec:discrepancy}
As discussed in detail in \citet{sallum_subm}, the discrepancy between the simulated and observed contrast curves most likely arises from the fact that the contrast is limited by the effect of charge migration \citep[the \textit{brighter-fatter} effect e.g.,][]{2020Hirata}, rather than being limited by the photon noise limit. The charge migration effect is exhibited by infrared detectors when the electric field induced by accumulated charges deflects new charges. This causes two nearby pixels to accumulate charge at different rates, with the brighter pixel apparently spilling photoelectrons into its neighbouring pixels. For brighter objects, this results in a  FWHM of the PSF with larger spatial extent. 

In addition to masking the presence of fainter companions in the close vicinity {(on the sky plane)} of a bright star, charge migration can cause brightness-dependent PSF differences (and thus calibration errors) between a science target and reference PSF target.
This effect was not taken into account during the generation of the simulated contrast curve shown in Figure \ref{fig:ContrastCurve}.
If we had the same level of charge migration between science and calibrator observations, we would have improved contrast quality and would have reached closer to the simulated achievable contrast \citep{sallum_subm}.
To plan observations with this mode in future cycles, observations should ideally target PSF references that well-matched to the science target in brightness.

\subsection{Mass Sensitivity Limits}
\label{ssec:masslimits}
The mass sensitivity accessible with the $5\sigma$ calculated contrast curve (see Section \ref{sec:ccurves}) was calculated using evolutionary models (see Figure \ref{fig:MassLimits}). An age value of $14\,\rm{Myr}$ \citep[based on the Lower Centaurus Crux age which HIP 65426 is a member of, as discussed in Appendix A of][]{2017Chauvin}, was used for the analysis across all used models. To account for uncertainties on the mass limit, different evolutionary models were used which encapsulate different physical processes. These are discussed in the following sections.

\begin{figure}
    \centering
    \includegraphics[scale=0.56]{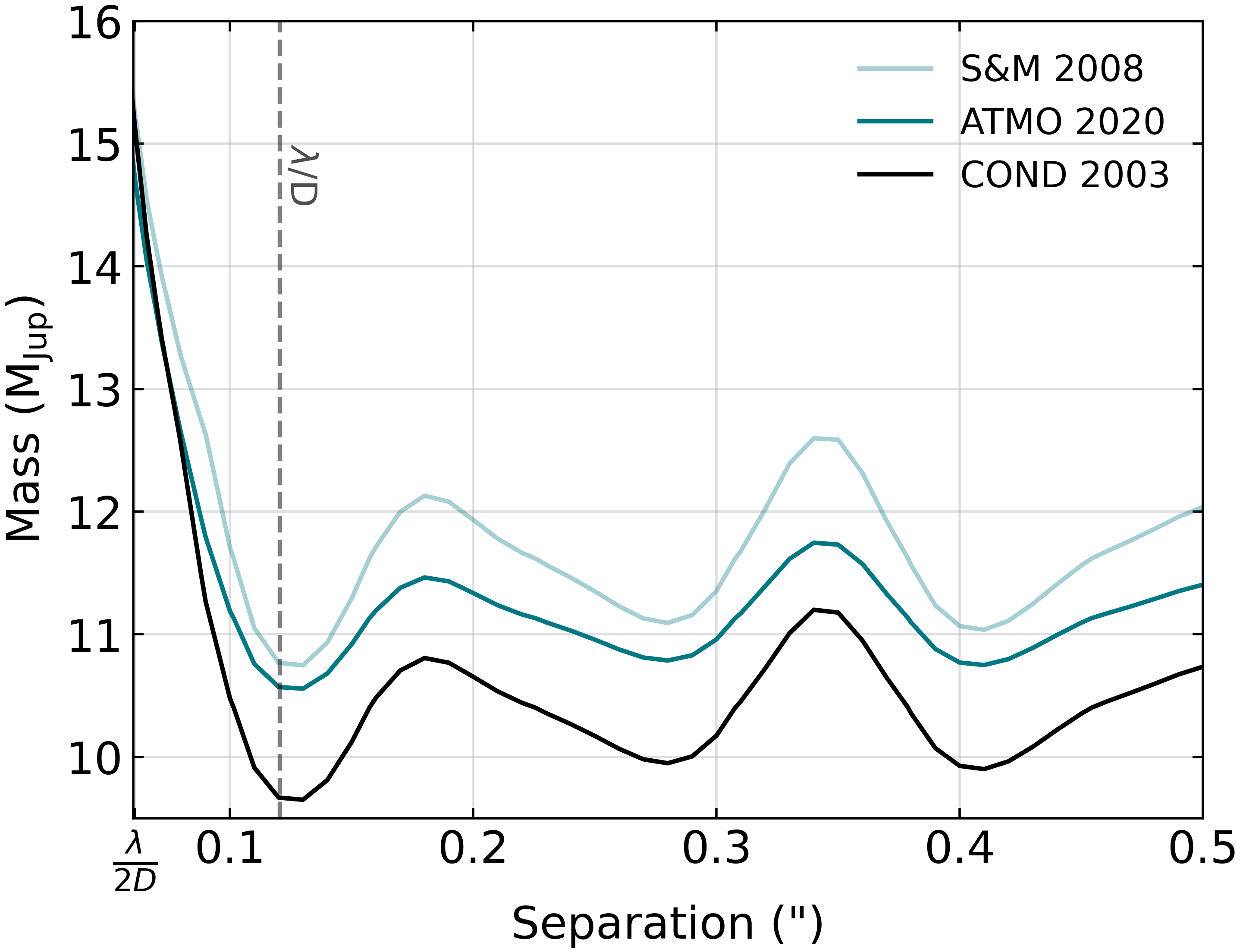}
    \caption{The accessible mass limits using the contrast curve from Figure \ref{fig:ContrastCurve} is shown in this Figure. The light cyan curve utilises the cloudy models from \cite{2008Saumon}, the dark cyan curve utilises the equilibrium case of \texttt{ATMO 2020} \citep{PhillipsPaper} and the black curve employs \texttt{COND 2003} \citep{2003Baraffe}. The lower limit on the x-axis is $0.5\lambda/D$, which is approximately the practical achievable inner working angle with AMI. The dashed grey line shows the conventional diffraction limit below which the contrast curve begins to grow steeply for accessible mass limits.}
    \label{fig:MassLimits}
\end{figure}

\subsubsection{\texttt{ATMO 2020} Atmospheric Models}
The mass detection limits were calculated from the generated contrast curve using the \texttt{ATMO} 2020 \citep{PhillipsPaper} set of models, similar to the method in \cite{2022Ray}. \texttt{ATMO} 2020 is a set of radiative-convective one-dimensional  equilibrium cloudless models describing the atmosphere and evolution of cool brown dwarfs and self-luminous giant exoplanets, spanning the mass range of  ${\sim}0.5{-}75\,\rm{M\textsubscript{Jup}}$. Even though these models are cloudless, they mimic the effect of clouds by using a lower temperature profile. The model is computed in three different sets of evolutionary models: one at
chemical equilibrium, and two at chemical disequilibrium assuming vertical mixing at different strengths. We keep our calculations and results limited to the case of equilibrium models, {since this case provides the baseline scenario of planetary atmospheric conditions, and does not take into account more complex considerations related to atmospheric dynamics, such as vertical atmospheric mixing \citep[][]{bmk11, 2013Konopacky,2023Currie}.} 

It is evident in Figure \ref{fig:MassLimits} that using \texttt{ATMO 2020}, mass values of ${\sim}11{-}12\,\rm{M\textsubscript{Jup}}$ are accessible at separations ${\sim}0.1{\arcsec}{-}0.4{\arcsec}$ or equivalently, ${\sim}1{-}3{\lambda /D}$ with our observations using the AMI mode. These separations are smaller than the inner working angles (IWAs) of the \textit{JWST} coronagraphs (${\sim}0.36{\arcsec}{-}0.50{\arcsec}$). This is achievable due to the combination of the interferometric capability of the AMI mode and the superior infrared sensitivity of \textit{JWST}.

The accessible mass values coincide with the deuterium burning mass limit for planetary mass companions \citep{2011Spiegel}. So, the \cite{2008Saumon} evolutionary models were also explored which takes into account the deuterium burning. This was done using the \texttt{species} toolkit \citep{2020Stolker} and is discussed in the following section.

\subsubsection{Saumon \& Marley 2008 and \texttt{COND 2003}\\Atmospheric Models}
The hybrid cloud grid from \cite{2008Saumon} was used to calculate the mass limits accessible with the given contrast in Figure \ref{fig:ContrastCurve}, as this model incorporates deuterium burning. Using this grid of evolutionary models ensures consistency with the \texttt{ATMO 2020} model (which mimics the effects of clouds) as well as the analysis and calculation of the bolometric flux of HIP\,65426\,b \citep{2023Carter}, the known companion in the system. Using this model, the mass limits accessible at separations ${\sim}0.1{\arcsec}{-}0.4{\arcsec}$ are ${\sim}11.0{-}12.5\,\rm{M\textsubscript{Jup}}$ (see Figure \ref{fig:MassLimits}). This hybrid cloud grid provides a simplified model of the  L–T transition by incorporating a cloudy atmosphere at high temperatures (T\textsubscript{eff}${\geq}1400\,\rm{K}$) and a cloud free atmosphere at low temperatures  (T\textsubscript{eff}${\leq}1200\,\rm{K}$). This grid is very similar to \texttt{COND 2003} \citep{2003Baraffe} and hence the latter was also explored to compute the mass limits in the absence of deuterium burning in the atmosphere of planetary mass companions. Using this model, mass limits accessible at separations ${\sim}0.1{\arcsec}{-}0.4{\arcsec}$ are ${\sim}9.5{-}11.0\,\rm{M\textsubscript{Jup}}$ (see Figure \ref{fig:MassLimits}).


\subsection{Mapping the probability of detecting companions}

\begin{figure*}
    \centering
    \includegraphics[scale=0.75]{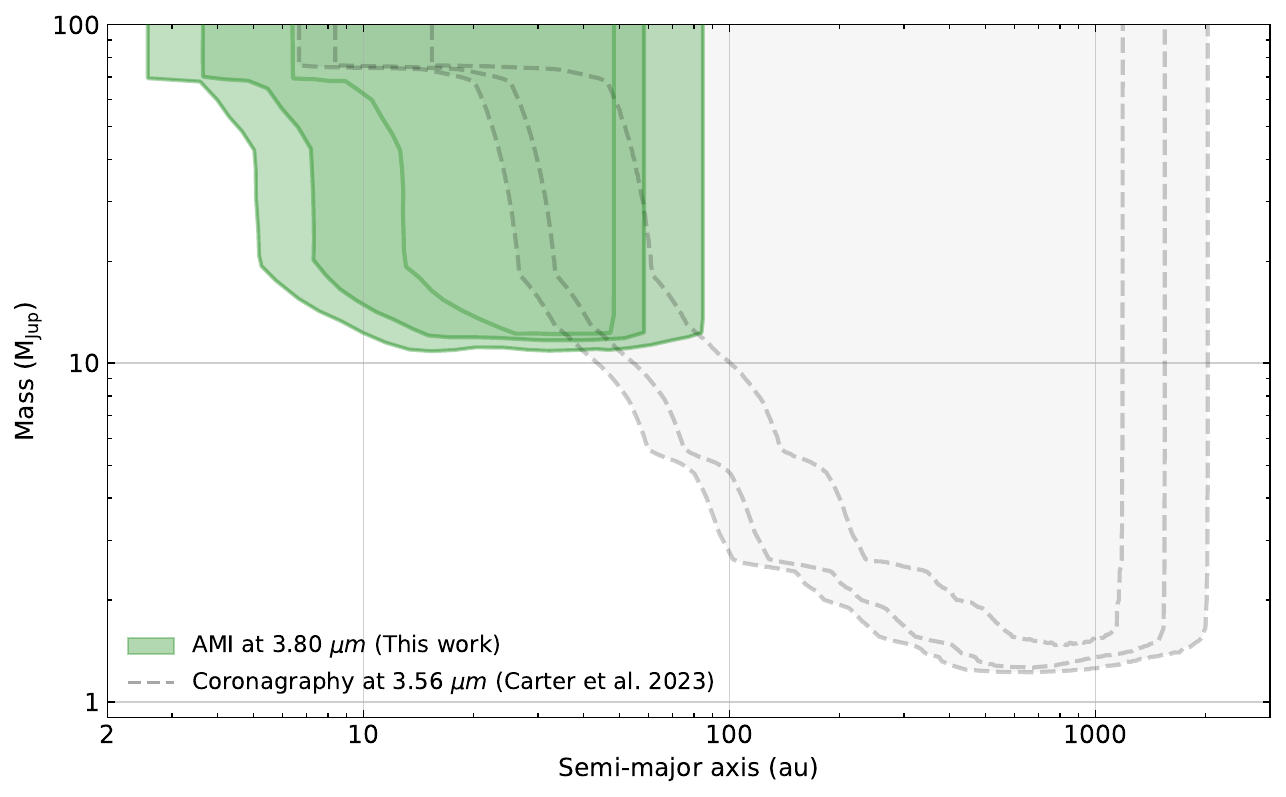}
    \caption{Detection probability maps (contours of 30\%, 68\% and 90\%) for \textit{JWST} observations, for the star HIP\,65426: green shows AMI in the F380M filter and the grey dashed lines show coronagraphy in the F356W filter. \textit{JWST}/AMI reaches close-in separations when compared to coronagraphy at similar wavelengths, probing an essentially unexplored parameter space.}
    \label{fig:DetProb}
\end{figure*}

The Exoplanet Detection Map Calculator \citep[\href{https://ascl.net/2010.008}{\texttt{Exo-DMC}},][]{ExoDMC} was used to estimate a detection probability map using the mass sensitivity limits. This tool uses a Monte Carlo approach to compare the instrument detection limits with a simulated, synthetic population of planets {with varying orbital geometries} around a given star to estimate the probability of detection of a companion of a given mass and semi-major axis. This information is then summarised in a detection probability map. This \textsc{Python} language tool is an adaptation of the previously existing code \texttt{MESS} \citep[Multi-purpose Exoplanet Simulation System,][]{MESS}. 

Figure \ref{fig:DetProb} shows the detection probability maps for the dataset described in this study (AMI at $3.80\,\rm{\mu m}$), as well as one for the coronagraphic observations of the same target obtained at a comparable wavelength ($3.56\,\rm{\mu m}$) described in \cite{2023Carter}. The figure clearly shows the exquisite capability of \textit{JWST}/NIRISS/AMI to detect companions at mid-infrared wavelengths in a completely new orbital parameter space. This can also be seen in Figure \ref{fig:SPHEREvsJWST} where \textit{JWST}/NIRISS/AMI probes separations lower than that of the \textit{VLT}/SPHERE/IRDIS observations at H-band \citep[$1.6\,\rm{\mu m}$,][]{2017Chauvin} for the same object. 

In this study we detect no additional companions but Figure \ref{fig:DetProb} clearly exhibits the capability of the AMI mode for the investigation of an unexplored parameter space of stellar systems with previously known companions. Figure 4 of \cite{2017Chauvin} shows that for this planetary system, regions ${\lesssim}20\,\rm{au}$, are inaccessible by \textit{VLT}/SPHERE. In Figure \ref{fig:DetProb}, we rule out the existence of any additional companions ${\gtrsim}10\,\rm{M\textsubscript{Jup}}$ at separations ${\sim}10{-}20\,\rm{au}$ around the host star. These observations hence provide sensitivity inside of the classical IWAs of \textit{JWST}'s conventional coronagraphs. 


\begin{figure}
    \centering
    \includegraphics[scale=0.53]{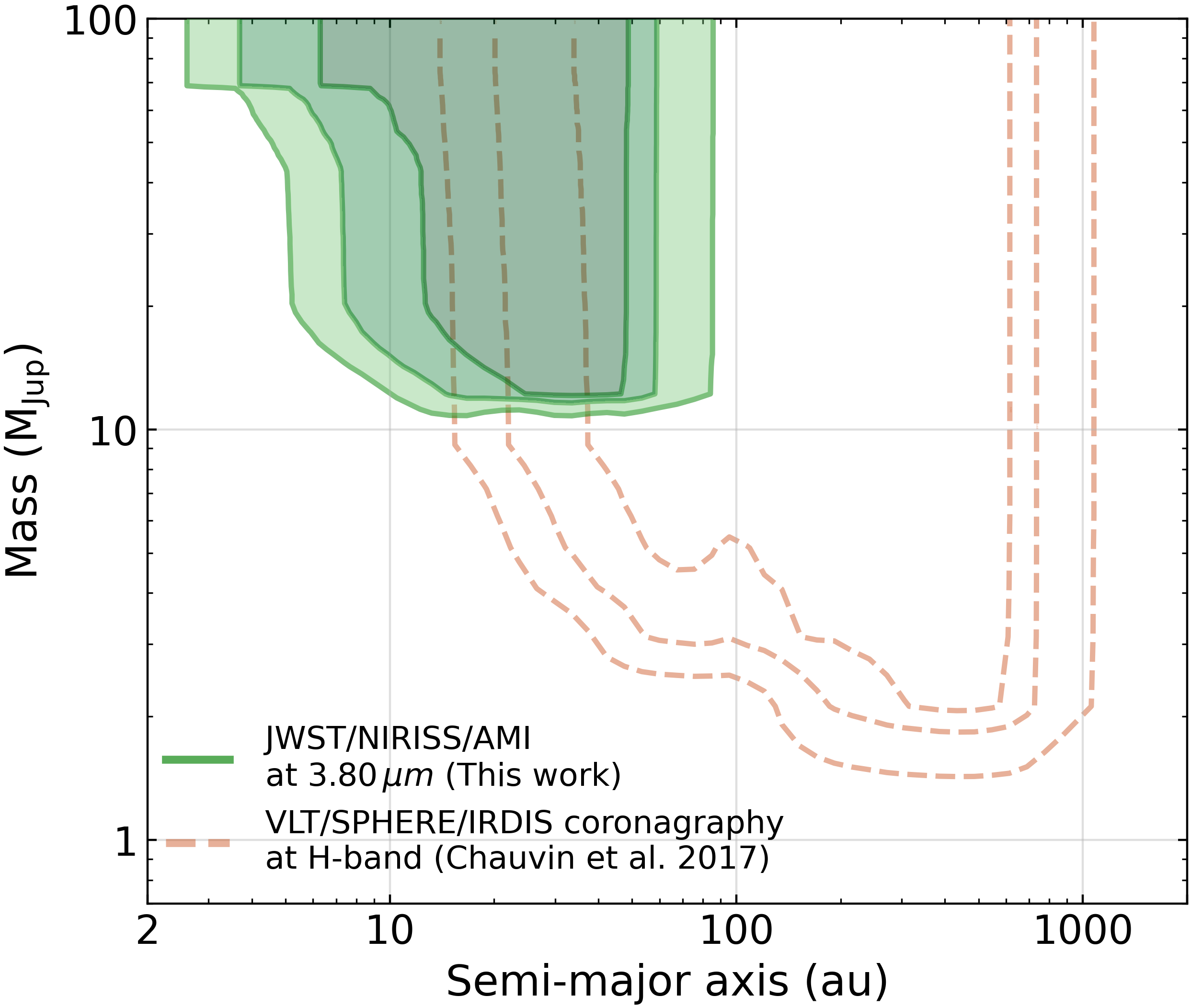}
    \caption{Detection probability maps (contours of 30\%, 68\% and, 90\%) for the AMI observation in F380M filter with \textit{JWST}/NIRISS, compared with the coronagraphic H\,band observations obtained with \textit{VLT}/SPHERE/IRDIS \citep{2017Chauvin}, for the star HIP\,65426.}
    \label{fig:SPHEREvsJWST}
\end{figure}

\section{Conclusions}
\label{sec:Conclusions}
In this study, we have demonstrated that the AMI mode with \textit{JWST}/NIRISS accesses a completely new parameter space (separations of ${\sim}0.05{\arcsec}{-}0.5{\arcsec}$), a region which is only accessible at lower contrast from the ground, and  largely inaccessible using the conventional coronagraphs on \textit{JWST}. This is evident from Figures \ref{fig:DetProb} and \ref{fig:SPHEREvsJWST} which exhibit that with \textit{JWST}/NIRISS/AMI, companions at Solar System planetary scales (as close in as ${\sim}10\,\rm{au}$) can be accessed. Solely in terms of contrast, the AMI mode on \textit{JWST} with a currently accessible {${5\sigma}$} 
 companion contrast of $\Delta m{\sim}7.5$, is still reaching roughly one magnitude deeper than very good ground-based AMI performance \citep[e.g. $\Delta m{\sim}6.5$ in][{calculated with a similar method}]{2023Vides}. This makes \textit{JWST}/AMI (even with the under-performance induced by charge migration), the most powerful single-telescope interferometer. And, with more robust reduction techniques and improved observation strategies this performance will potentially improve in the future.

Going forward, this mode will be the prime technique to detect companions around stars in the closest star forming regions at close-in separations. The number of stars in young moving groups available for planet searches is limited to ${\sim}100$ targets \citep[for example in the moving groups of $\beta$ Pictoris and TW Hydrae,][]{2022Ray}. However, targeting stars in the Taurus-Auriga or Scorpius-Centaurus associations (which HIP\,65426 is a member of), increases the number of such available targets by ${\sim}1{-}2$ orders of magnitude since these associations are potentially rich in thousands of such targets. Opening up the possibility to probe the members in these star formation regions would mean that many promising targets (e.g., those with debris disks, evidence for accretion or protoplanets using ALMA gas kinematics, etc.) can be probed. These targets could be part of future \textit{JWST} (and \textit{ELT}) observations.

This mode can also be used to find planets in systems which have previously known companions. The characterisation of the multiplicity of PMCs around nearby stars would be incomplete without probing inner regions, which \textit{JWST}/NIRISS/AMI can access. Previous studies \citep{2010Marois,Wagner:2019ApJ,2020Nowak,2023Hinkley} have shown that additional companions at close in separations can be found in systems that already have a known companion. This is beyond the capabilities of conventional coronagraphs on \textit{JWST} due to their relatively poor IWAs. Hence observations in future cycles using \textit{JWST}/NIRISS/AMI will shed some light on this unexplored piece of parameter space.

The angular parameter space (${\sim}100{-}300\,\rm{mas}$, see Figure \ref{fig:ContrastCurve}) accessible with \textit{JWST/AMI} for the case of members in nearby stellar associations, overlaps the peak sensitivity of current and future ground-based interferometers such as GRAVITY \citep{2019Lacour,2020Nowak} and BiFROST \citep{2022Kraus}. These interferometers can measure dynamical masses of PMCs with high precision (up to ${\sim10\%}$) over a short portion of their orbit \citep[e.g.][]{2023Hinkley}. The combination of the measurements of precise dynamical masses and the PMC brightness at ${\sim}3{-}5\,\rm{\mu m}$ (near the peak of their SED, which gives tightly constrained measurements of bolometric luminosity) is exceedingly powerful for constraining atmospheric and evolutionary models that are highly uncertain at young ages \citep[see Figure 11,][]{2022Ray}.
In addition to this, the obtained contrast limits (in Figures \ref{fig:ContrastCurve} and \ref{fig:MassLimits}) will also access the expected luminosities of circumplanetary disks whose spectral energy distributions can be used to inform planet formation timescales and exomoon formation. 
Through these applications and more, NIRISS AMI observations in future cycles will provide new, direct constraints on planet formation and evolution.

\section*{Acknowledgements}
This work is based on observations made with the NASA/ESA/CSA \textit{JWST} and were obtained from the Mikulski Archive for Space Telescopes (MAST) at the Space Telescope Science Institute. The specific observations analyzed can be accessed via\dataset[DOI: 10.17909/8by2-x206]{https://doi.org/10.17909/8by2-x206}. We are truly grateful for the countless hours that thousands of people have devoted
to the design, construction, and commissioning of \textit{JWST}. We thank the anonymous referee for their comments which have been crucial {towards the improvement of this paper}. This project was supported by a grant from STScI (JWST-ERS-01386) under NASA contract NAS5-03127. SR was supported by the Global Excellence Award by the University of Exeter. This work is based in part on observations obtained at the Southern Astrophysical Research (\textit{SOAR}) telescope, which is a joint project of the Minist\'{e}rio da Ci\^{e}ncia, Tecnologia e Inova\c{c}\~{o}es (MCTI/LNA) do Brasil, the US National Science Foundation’s NOIRLab, the University of North Carolina
at Chapel Hill (UNC), and Michigan State University (MSU). This work has also made use of the SPHERE Data Centre, jointly operated by OSUG/IPAG (Grenoble), PYTHEAS/LAM/CeSAM (Marseille), OCA/Lagrange (Nice), Observatoire de Paris/LESIA (Paris), and Observatoire de Lyon/CRAL, and is supported by a grant from Labex OSUG@2020 (Investissements d’avenir – ANR10 LABX56). This work benefited from the 2022 Exoplanet
Summer Program in the Other Worlds Laboratory (OWL) at the
University of California, Santa Cruz, a program funded by the
Heising-Simons Foundation.


\bibliography{sample631,MasterBiblio_Sasha.bib,example.bib}

\begin{thebibliography}{}
\expandafter\ifx\csname natexlab\endcsname\relax\def\natexlab#1{#1}\fi
\providecommand{\url}[1]{\href{#1}{#1}}
\providecommand{\dodoi}[1]{doi:~\href{http://doi.org/#1}{\nolinkurl{#1}}}
\providecommand{\doeprint}[1]{\href{http://ascl.net/#1}{\nolinkurl{http://ascl.net/#1}}}
\providecommand{\doarXiv}[1]{\href{https://arxiv.org/abs/#1}{\nolinkurl{https://arxiv.org/abs/#1}}}

\bibitem[{{Baldwin} {et~al.}(1986){Baldwin}, {Haniff}, {Mackay}, \& {Warner}}]{1986Baldwin}
{Baldwin}, J.~E., {Haniff}, C.~A., {Mackay}, C.~D., \& {Warner}, P.~J. 1986, \nat, 320, 595, \dodoi{10.1038/320595a0}

\bibitem[{{Baraffe} {et~al.}(2003){Baraffe}, {Chabrier}, {Barman}, {Allard}, \& {Hauschildt}}]{2003Baraffe}
{Baraffe}, I., {Chabrier}, G., {Barman}, T.~S., {Allard}, F., \& {Hauschildt}, P.~H. 2003, \aap, 402, 701, \dodoi{10.1051/0004-6361:20030252}

\bibitem[{{Barman} {et~al.}(2011){Barman}, {Macintosh}, {Konopacky}, \& {Marois}}]{bmk11}
{Barman}, T.~S., {Macintosh}, B., {Konopacky}, Q.~M., \& {Marois}, C. 2011, \apj, 733, 65, \dodoi{10.1088/0004-637X/733/1/65}

\bibitem[{{Bonavita}(2020)}]{ExoDMC}
{Bonavita}, M. 2020, {Exo-DMC: Exoplanet Detection Map Calculator}.
\newblock \doeprint{2010.008}

\bibitem[{{Bonavita} {et~al.}(2012){Bonavita}, {Chauvin}, {Desidera}, {Gratton}, {Janson}, {Beuzit}, {Kasper}, \& {Mordasini}}]{MESS}
{Bonavita}, M., {Chauvin}, G., {Desidera}, S., {et~al.} 2012, \aap, 537, A67, \dodoi{10.1051/0004-6361/201116852}

\bibitem[{{Bonneau} {et~al.}(2006){Bonneau}, {Clausse}, {Delfosse}, {Mourard}, {Cetre}, {Chelli}, {Cruzal{\`e}bes}, {Duvert}, \& {Zins}}]{2006Bonneau}
{Bonneau}, D., {Clausse}, J.~M., {Delfosse}, X., {et~al.} 2006, \aap, 456, 789, \dodoi{10.1051/0004-6361:20054469}

\bibitem[{{Bushouse} {et~al.}(2022){Bushouse}, {Eisenhamer}, {Dencheva}, {Davies}, {Greenfield}, {Morrison}, {Hodge}, {Simon}, {Grumm}, {Droettboom}, {Slavich}, {Sosey}, {Pauly}, {Miller}, {Jedrzejewski}, {Hack}, {Davis}, {Crawford}, {Law}, {Gordon}, {Regan}, {Cara}, {MacDonald}, {Bradley}, {Shanahan}, \& {Jamieson}}]{Bushouse2022}
{Bushouse}, H., {Eisenhamer}, J., {Dencheva}, N., {et~al.} 2022, {spacetelescope/jwst: JWST 1.6.2}, 1.6.2, Zenodo,  Zenodo, \dodoi{10.5281/zenodo.6984366}

\bibitem[{{Carter} {et~al.}(2023){Carter}, {Hinkley}, {Kammerer}, {Skemer}, {Biller}, {Leisenring}, {Millar-Blanchaer}, {Petrus}, {Stone}, {Ward-Duong}, {Wang}, {Girard}, {Hines}, {Perrin}, {Pueyo}, {Balmer}, {Bonavita}, {Bonnefoy}, {Chauvin}, {Choquet}, {Christiaens}, {Danielski}, {Kennedy}, {Matthews}, {Miles}, {Patapis}, {Ray}, {Rickman}, {Sallum}, {Stapelfeldt}, {Whiteford}, {Zhou}, {Absil}, {Boccaletti}, {Booth}, {Bowler}, {Chen}, {Currie}, {Fortney}, {Grady}, {Greebaum}, {Henning}, {Hoch}, {Janson}, {Kalas}, {Kenworthy}, {Kervella}, {Kraus}, {Lagage}, {Liu}, {Macintosh}, {Marino}, {Marley}, {Marois}, {Matthews}, {Mawet}, {McElwain}, {Metchev}, {Meyer}, {Molliere}, {Moran}, {Morley}, {Mukherjee}, {Pantin}, {Quirrenbach}, {Rebollido}, {Ren}, {Schneider}, {Vasist}, {Worthen}, {Wyatt}, {Briesemeister}, {Bryan}, {Calissendorff}, {Cantalloube}, {Cugno}, {De Furio}, {Dupuy}, {Factor}, {Faherty}, {Fitzgerald}, {Franson}, {Gonzales}, {Hood}, {Howe}, {Kuzuhara}, {Lagrange}, {Lawson}, {Lazzoni}, {Lew}, {Liu},
  {Llop-Sayson}, {Lloyd}, {Martinez}, {Mazoyer}, {Palma-Bifani}, {Quanz}, {Redai}, {Samland}, {Schlieder}, {Tamura}, {Tan}, {Uyama}, {Vigan}, {Vos}, {Wagner}, {Wolff}, {Ygouf}, {Zhang}, {Zhang}, \& {Zhang}}]{2023Carter}
{Carter}, A.~L., {Hinkley}, S., {Kammerer}, J., {et~al.} 2023, \apjl, 951, L20, \dodoi{10.3847/2041-8213/acd93e}

\bibitem[{{Chauvin} {et~al.}(2017){Chauvin}, {Desidera}, {Lagrange}, {Vigan}, {Gratton}, {Langlois}, {Bonnefoy}, {Beuzit}, {Feldt}, {Mouillet}, {Meyer}, {Cheetham}, {Biller}, {Boccaletti}, {D'Orazi}, {Galicher}, {Hagelberg}, {Maire}, {Mesa}, {Olofsson}, {Samland}, {Schmidt}, {Sissa}, {Bonavita}, {Charnay}, {Cudel}, {Daemgen}, {Delorme}, {Janin-Potiron}, {Janson}, {Keppler}, {Le Coroller}, {Ligi}, {Marleau}, {Messina}, {Molli{\`e}re}, {Mordasini}, {M{\"u}ller}, {Peretti}, {Perrot}, {Rodet}, {Rouan}, {Zurlo}, {Dominik}, {Henning}, {Menard}, {Schmid}, {Turatto}, {Udry}, {Vakili}, {Abe}, {Antichi}, {Baruffolo}, {Baudoz}, {Baudrand}, {Blanchard}, {Bazzon}, {Buey}, {Carbillet}, {Carle}, {Charton}, {Cascone}, {Claudi}, {Costille}, {Deboulbe}, {De Caprio}, {Dohlen}, {Fantinel}, {Feautrier}, {Fusco}, {Gigan}, {Giro}, {Gisler}, {Gluck}, {Hubin}, {Hugot}, {Jaquet}, {Kasper}, {Madec}, {Magnard}, {Martinez}, {Maurel}, {Le Mignant}, {M{\"o}ller-Nilsson}, {Llored}, {Moulin}, {Orign{\'e}}, {Pavlov}, {Perret}, {Petit},
  {Pragt}, {Puget}, {Rabou}, {Ramos}, {Rigal}, {Rochat}, {Roelfsema}, {Rousset}, {Roux}, {Salasnich}, {Sauvage}, {Sevin}, {Soenke}, {Stadler}, {Suarez}, {Weber}, {Wildi}, {Antoniucci}, {Augereau}, {Baudino}, {Brandner}, {Engler}, {Girard}, {Gry}, {Kral}, {Kopytova}, {Lagadec}, {Milli}, {Moutou}, {Schlieder}, {Szul{\'a}gyi}, {Thalmann}, \& {Wahhaj}}]{2017Chauvin}
{Chauvin}, G., {Desidera}, S., {Lagrange}, A.~M., {et~al.} 2017, \aap, 605, L9, \dodoi{10.1051/0004-6361/201731152}

\bibitem[{{Cheetham} {et~al.}(2016){Cheetham}, {Girard}, {Lacour}, {Schworer}, {Haubois}, \& {Beuzit}}]{2016Cheetham}
{Cheetham}, A.~C., {Girard}, J., {Lacour}, S., {et~al.} 2016, in Society of Photo-Optical Instrumentation Engineers (SPIE) Conference Series, Vol. 9907, Optical and Infrared Interferometry and Imaging V, ed. F.~{Malbet}, M.~J. {Creech-Eakman}, \& P.~G. {Tuthill}, 99072T, \dodoi{10.1117/12.2231983}

\bibitem[{{Cheetham} {et~al.}(2019){Cheetham}, {Samland}, {Brems}, {Launhardt}, {Chauvin}, {S{\'e}gransan}, {Henning}, {Quirrenbach}, {Avenhaus}, {Cugno}, {Girard}, {Godoy}, {Kennedy}, {Maire}, {Metchev}, {M{\"u}ller}, {Musso Barcucci}, {Olofsson}, {Pepe}, {Quanz}, {Queloz}, {Reffert}, {Rickman}, {van Boekel}, {Boccaletti}, {Bonnefoy}, {Cantalloube}, {Charnay}, {Delorme}, {Janson}, {Keppler}, {Lagrange}, {Langlois}, {Lazzoni}, {Menard}, {Mesa}, {Meyer}, {Schmidt}, {Sissa}, {Udry}, \& {Zurlo}}]{2019Cheetham}
{Cheetham}, A.~C., {Samland}, M., {Brems}, S.~S., {et~al.} 2019, \aap, 622, A80, \dodoi{10.1051/0004-6361/201834112}

\bibitem[{{Currie} {et~al.}(2023){Currie}, {Brandt}, {Brandt}, {Lacy}, {Burrows}, {Guyon}, {Tamura}, {Liu}, {Sagynbayeva}, {Tobin}, {Chilcote}, {Groff}, {Marois}, {Thompson}, {Murphy}, {Kuzuhara}, {Lawson}, {Lozi}, {Deo}, {Vievard}, {Skaf}, {Uyama}, {Jovanovic}, {Martinache}, {Kasdin}, {Kudo}, {McElwain}, {Janson}, {Wisniewski}, {Hodapp}, {Nishikawa}, {He{\l}miniak}, {Kwon}, \& {Hayashi}}]{2023Currie}
{Currie}, T., {Brandt}, G.~M., {Brandt}, T.~D., {et~al.} 2023, Science, 380, 198, \dodoi{10.1126/science.abo6192}

\bibitem[{{Doyon} {et~al.}(2012){Doyon}, {Hutchings}, {Beaulieu}, {Albert}, {Lafreni{\`e}re}, {Willott}, {Touahri}, {Rowlands}, {Maszkiewicz}, {Fullerton}, {Volk}, {Martel}, {Chayer}, {Sivaramakrishnan}, {Abraham}, {Ferrarese}, {Jayawardhana}, {Johnstone}, {Meyer}, {Pipher}, \& {Sawicki}}]{DoyonNIRISS}
{Doyon}, R., {Hutchings}, J.~B., {Beaulieu}, M., {et~al.} 2012, in Society of Photo-Optical Instrumentation Engineers (SPIE) Conference Series, Vol. 8442, Space Telescopes and Instrumentation 2012: Optical, Infrared, and Millimeter Wave, ed. M.~C. {Clampin}, G.~G. {Fazio}, H.~A. {MacEwen}, \& J.~{Oschmann}, Jacobus~M., 84422R, \dodoi{10.1117/12.926578}

\bibitem[{{Doyon} {et~al.}(2023){Doyon}, {Willott}, {Hutchings}, {Sivaramakrishnan}, {Albert}, {Lafreniere}, {Rowlands}, {Begona Vila}, {Martel}, {LaMassa}, {Aldridge}, {Artigau}, {Cameron}, {Chayer}, {Cook}, {Cooper}, {Darveau-Bernier}, {Dupuis}, {Earnshaw}, {Espinoza}, {Filippazzo}, {Fullerton}, {Gaudreau}, {Gawlik}, {Goudfrooij}, {Haley}, {Kammerer}, {Kendall}, {Lambros}, {Ilinca Ignat}, {Maszkiewicz}, {McColgan}, {Morishita}, {Ouellette}, {Pacifici}, {Philippi}, {Radica}, {Ravindranath}, {Rowe}, {Roy}, {Saad}, {Sohn}, {Talens}, {Thatte}, {Taylor}, {Vandal}, {Volk}, {Wander}, {Warner}, {Zheng}, {Zhou}, {Abraham}, {Beaulieu}, {Benneke}, {Ferrarese}, {Johnstone}, {Kaltenegger}, {Meyer}, {Pipher}, {Rameau}, {Rieke}, {Salhi}, \& {Sawicki}}]{2023Doyon}
{Doyon}, R., {Willott}, C.~J., {Hutchings}, J.~B., {et~al.} 2023, arXiv e-prints, arXiv:2306.03277, \dodoi{10.48550/arXiv.2306.03277}

\bibitem[{{Foreman-Mackey} {et~al.}(2013){Foreman-Mackey}, {Hogg}, {Lang}, \& {Goodman}}]{2013_Foreman-Mackey}
{Foreman-Mackey}, D., {Hogg}, D.~W., {Lang}, D., \& {Goodman}, J. 2013, \pasp, 125, 306, \dodoi{10.1086/670067}

\bibitem[{{Gardner} {et~al.}(2006){Gardner}, {Mather}, {Clampin}, {Doyon}, {Greenhouse}, {Hammel}, {Hutchings}, {Jakobsen}, {Lilly}, {Long}, {Lunine}, {McCaughrean}, {Mountain}, {Nella}, {Rieke}, {Rieke}, {Rix}, {Smith}, {Sonneborn}, {Stiavelli}, {Stockman}, {Windhorst}, \& {Wright}}]{2006Gardner}
{Gardner}, J.~P., {Mather}, J.~C., {Clampin}, M., {et~al.} 2006, \ssr, 123, 485, \dodoi{10.1007/s11214-006-8315-7}

\bibitem[{{Gardner} {et~al.}(2023){Gardner}, {Mather}, {Abbott}, {Abell}, {Abernathy}, {Abney}, {Abraham}, {Abraham}, {Abul-Huda}, {Acton}, {Adams}, {Adams}, {Adler}, {Adriaensen}, {Aguilar}, {Ahmed}, {Ahmed}, {Ahmed}, {Albat}, {Albert}, {Alberts}, {Aldridge}, {Allen}, {Allen}, {Altenburg}, {Altunc}, {Alvarez}, {{\'A}lvarez-M{\'a}rquez}, {Alves de Oliveira}, {Ambrose}, {Anandakrishnan}, {Andersen}, {Anderson}, {Anderson}, {Anderson}, {Anderson}, {Aprea}, {Archer}, {Arenberg}, {Argyriou}, {Arribas}, {Artigau}, {Arvai}, {Atcheson}, {Atkinson}, {Averbukh}, {Aymergen}, {Bacinski}, {Baggett}, {Bagnasco}, {Baker}, {Balzano}, {Banks}, {Baran}, {Barker}, {Barrett}, {Barringer}, {Barto}, {Bast}, {Baudoz}, {Baum}, {Beatty}, {Beaulieu}, {Bechtold}, {Beck}, {Beddard}, {Beichman}, {Bellagama}, {Bely}, {Berger}, {Bergeron}, {Bernier}, {Bertch}, {Beskow}, {Betz}, {Biagetti}, {Birkmann}, {Bjorklund}, {Blackwood}, {Blazek}, {Blossfeld}, {Bluth}, {Boccaletti}, {Boegner}, {Bohlin}, {Boia}, {B{\"o}ker}, {Bonaventura}, {Bond},
  {Bosley}, {Boucarut}, {Bouchet}, {Bouwman}, {Bower}, {Bowers}, {Bowers}, {Boyce}, {Boyer}, {Boyer}, {Boyer}, {Boyer}, {Bradley}, {Brady}, {Brandl}, {Brannen}, {Breda}, {Bremmer}, {Brennan}, {Bresnahan}, {Bright}, {Broiles}, {Bromenschenkel}, {Brooks}, {Brooks}, {Brown}, {Brown}, {Brown}, {Bruce}, {Bryson}, {Bujanda}, {Bullock}, {Bunker}, {Bureo}, {Burt}, {Bush}, {Bushouse}, {Bussman}, {Cabaud}, {Cale}, {Calhoon}, {Calvani}, {Canipe}, {Caputo}, {Cara}, {Carey}, {Case}, {Cesari}, {Cetorelli}, {Chance}, {Chandler}, {Chaney}, {Chapman}, {Charlot}, {Chayer}, {Cheezum}, {Chen}, {Chen}, {Cherinka}, {Chichester}, {Chilton}, {Chittiraibalan}, {Clampin}, {Clark}, {Clark}, {Clark}, {Claybrooks}, {Cleveland}, {Cohen}, {Cohen}, {Col{\'o}n}, {Coleman}, {Colina}, {Comber}, {Comeau}, {Comer}, {Conde Reis}, {Connolly}, {Conroy}, {Contos}, {Contreras}, {Cook}, {Cooper}, {Cooper}, {Correia}, {Correnti}, {Cossou}, {Costanza}, {Coulais}, {Cox}, {Coyle}, {Cracraft}, {Crew}, {Curtis}, {Cusveller}, {Da Costa Maciel}, {Dailey},
  {Daugeron}, {Davidson}, {Davies}, {Davis}, {Davis}, {Day}, {de Chambure}, {de Jong}, {De Marchi}, {Dean}, {Decker}, {Delisa}, {Dell}, {Dellagatta}, {Dembinska}, {Demosthenes}, {Dencheva}, {Deneu}, {DePriest}, {Deschenes}, {Dethienne}, {Detre}, {Diaz}, {Dicken}, {DiFelice}, {Dillman}, {Disharoon}, {Dixon}, {Doggett}, {Dominguez}, {Donaldson}, {Doria-Warner}, {Santos}, {Doty}, {Douglas}, {Doyon}, {Dressler}, {Driggers}, {Driggers}, {Dunn}, {DuPrie}, {Dupuis}, {Durning}, {Dutta}, {Earl}, {Eccleston}, {Ecobichon}, {Egami}, {Ehrenwinkler}, {Eisenhamer}, {Eisenhower}, {Eisenstein}, {El Hamel}, {Elie}, {Elliott}, {Elliott}, {Engesser}, {Espinoza}, {Etienne}, {Etxaluze}, {Evans}, {Fabreguettes}, {Falcolini}, {Falini}, {Fatig}, {Feeney}, {Feinberg}, {Fels}, {Ferdous}, {Ferguson}, {Ferrarese}, {Ferreira}, {Ferruit}, {Ferry}, {Filippazzo}, {Firre}, {Fix}, {Flagey}, {Flanagan}, {Fleming}, {Florian}, {Flynn}, {Foiadelli}, {Fontaine}, {Fontanella}, {Forshay}, {Fortner}, {Fox}, {Framarini}, {Francisco}, {Franck}, {Franx},
  {Franz}, {Friedman}, {Friend}, {Frost}, {Fu}, {Fullerton}, {Gaillard}, {Galkin}, {Gallagher}, {Galyer}, {Garc{\'\i}a Mar{\'\i}n}, {Gardner}, {Garland}, {Garrett}, {Gasman}, {G{\'a}sp{\'a}r}, {Gastaud}, {Gaudreau}, {Gauthier}, {Geers}, {Geithner}, {Gennaro}, {Gerber}, {Gereau}, {Giampaoli}, {Giardino}, {Gibbons}, {Gilbert}, {Gilman}, {Girard}, {Giuliano}, {Gkountis}, {Glasse}, {Glassmire}, {Glauser}, {Glazer}, {Goldberg}, {Golimowski}, {Gonzaga}, {Gordon}, {Gordon}, {Goudfrooij}, {Gough}, {Graham}, {Grau}, {Green}, {Greene}, {Greene}, {Greenfield}, {Greenhouse}, {Greve}, {Greville}, {Grimaldi}, {Groe}, {Groebner}, {Grumm}, {Grundy}, {G{\"u}del}, {Guillard}, {Guldalian}, {Gunn}, {Gurule}, {Gutman}, {Guy}, {Guyot}, {Hack}, {Haderlein}, {Hagan}, {Hagedorn}, {Hainline}, {Haley}, {Hami}, {Hamilton}, {Hammann}, {Hammel}, {Hanley}, {Hansen}, {Hardy}, {Harnisch}, {Harr}, {Harris}, {Hart}, {Hartig}, {Hasan}, {Hashim}, {Hashimoto}, {Haskins}, {Hawkins}, {Hayden}, {Hayden}, {Healy}, {Hecht}, {Heeg}, {Hejal}, {Helm},
  {Hengemihle}, {Henning}, {Henry}, {Henry}, {Henshaw}, {Hernandez}, {Herrington}, {Heske}, {Hesman}, {Hickey}, {Hilbert}, {Hines}, {Hinz}, {Hirsch}, {Hitcho}, {Hodapp}, {Hodge}, {Hoffman}, {Holfeltz}, {Holler}, {Hoppa}, {Horner}, {Howard}, {Howard}, {Huber}, {Hunkeler}, {Hunter}, {Hunter}, {Hurd}, {Hurst}, {Hutchings}, {Hylan}, {Ignat}, {Illingworth}, {Irish}, {Isaacs}, {Jackson}, {Jaffe}, {Jahic}, {Jahromi}, {Jakobsen}, {James}, {James}, {James}, {Jamieson}, {Jandra}, {Jayawardhana}, {Jedrzejewski}, {Jeffers}, {Jensen}, {Joanne}, {Johns}, {Johnson}, {Johnson}, {Johnson}, {Johnson}, {Johnson}, {Johnson}, {Johnstone}, {Jollet}, {Jones}, {Jones}, {Jones}, {Jones}, {Jones}, {Jordan}, {Jordan}, {Jue}, {Jurkowski}, {Justis}, {Justtanont}, {Kaleida}, {Kalirai}, {Kalmanson}, {Kaltenegger}, {Kammerer}, {Kan}, {Kanarek}, {Kao}, {Karakla}, {Karl}, {Kassin}, {Kauffman}, {Kavanagh}, {Kelley}, {Kelly}, {Kendrew}, {Kennedy}, {Kenny}, {Keski-Kuha}, {Keyes}, {Khan}, {Kidwell}, {Kimble}, {King}, {King}, {Kinzel}, {Kirk},
  {Kirkpatrick}, {Klaassen}, {Klingemann}, {Klintworth}, {Knapp}, {Knight}, {Knollenberg}, {Knutsen}, {Koehler}, {Koekemoer}, {Kofler}, {Kontson}, {Kovacs}, {Kozhurina-Platais}, {Krause}, {Kriss}, {Krist}, {Kristoffersen}, {Krogel}, {Krueger}, {Kulp}, {Kumari}, {Kwan}, {Kyprianou}, {Labador}, {Labiano}, {Lafreni{\`e}re}, {Lagage}, {Laidler}, {Laine}, {Laird}, {Lajoie}, {Lallo}, {Lam}, {LaMassa}, {Lambros}, {Lampenfield}, {Lander}, {Langston}, {Larson}, {Larson}, {LaVerghetta}, {Law}, {Lawrence}, {Lee}, {Lee}, {Lee}, {Leisenring}, {Leveille}, {Levenson}, {Levi}, {Levine}, {Lewis}, {Lewis}, {Lewis}, {Libralato}, {Lidon}, {Liebrecht}, {Lightsey}, {Lilly}, {Lim}, {Lim}, {Ling}, {Link}, {Link}, {Lipinski}, {Liu}, {Lo}, {Lobmeyer}, {Logue}, {Long}, {Long}, {Long}, {Long}, {L{\'o}pez-Caniego}, {Lotz}, {Love-Pruitt}, {Lubskiy}, {Luers}, {Luetgens}, {Luevano}, {Lui}, {Lund}, {Lundquist}, {Lunine}, {L{\"u}tzgendorf}, {Lynch}, {MacDonald}, {MacDonald}, {Macias}, {Macklis}, {Maghami}, {Maharaja}, {Maiolino},
  {Makrygiannis}, {Malla}, {Malumuth}, {Manjavacas}, {Marini}, {Marrione}, {Marston}, {Martel}, {Martin}, {Martin}, {Martinez}, {Maschmann}, {Masci}, {Masetti}, {Maszkiewicz}, {Matthews}, {Matuskey}, {McBrayer}, {McCarthy}, {McCaughrean}, {McClare}, {McClare}, {McCloskey}, {McClurg}, {McCoy}, {McElwain}, {McGregor}, {McGuffey}, {McKay}, {McKenzie}, {McLean}, {McMaster}, {McNeil}, {De Meester}, {Mehalick}, {Meixner}, {Mel{\'e}ndez}, {Menzel}, {Menzel}, {Merz}, {Mesterharm}, {Meyer}, {Meyett}, {Meza}, {Midwinter}, {Milam}, {Miller}, {Miller}, {Miskey}, {Misselt}, {Mitchell}, {Mohan}, {Montoya}, {Moran}, {Morishita}, {Moro-Mart{\'\i}n}, {Morrison}, {Morrison}, {Morse}, {Moschos}, {Moseley}, {Mosier}, {Mosner}, {Mountain}, {Muckenthaler}, {Mueller}, {Mueller}, {Muhiem}, {M{\"u}hlmann}, {Mullally}, {Mullen}, {Munger}, {Murphy}, {Murray}, {Muzerolle}, {Mycroft}, {Myers}, {Myers}, {Myers}, {Myers}, {Myrick}, {Nagle}, {Nayak}, {Naylor}, {Neff}, {Nelan}, {Nella}, {Nguyen}, {Nguyen}, {Nickson}, {Nidhiry}, {Niedner},
  {Nieto-Santisteban}, {Nikolov}, {Nishisaka}, {Noriega-Crespo}, {Nota}, {O'Mara}, {Oboryshko}, {O'Brien}, {Ochs}, {Offenberg}, {Ogle}, {Ohl}, {Olmsted}, {Osborne}, {O'Shaughnessy}, {{\"O}stlin}, {O'Sullivan}, {Otor}, {Ottens}, {Ouellette}, {Outlaw}, {Owens}, {Pacifici}, {Page}, {Paranilam}, {Park}, {Parrish}, {Paschal}, {Patapis}, {Patel}, {Patrick}, {Pattishall}, {Paul}, {Paul}, {Pauly}, {Pavlovsky}, {Pe{\~n}a-Guerrero}, {Pedder}, {Peek}, {Pelham}, {Penanen}, {Perriello}, {Perrin}, {Perrine}, {Perrygo}, {Peslier}, {Petach}, {Peterson}, {Pfarr}, {Pierson}, {Pietraszkiewicz}, {Pilchen}, {Pipher}, {Pirzkal}, {Pitman}, {Player}, {Plesha}, {Plitzke}, {Pohner}, {Poletis}, {Pollizzi}, {Polster}, {Pontius}, {Pontoppidan}, {Porges}, {Potter}, {Prescott}, {Proffitt}, {Pueyo}, {Quispe Neira}, {Radich}, {Rager}, {Rameau}, {Ramey}, {Ramos Alarcon}, {Rampini}, {Rapp}, {Rashford}, {Rauscher}, {Ravindranath}, {Rawle}, {Rawlings}, {Ray}, {Regan}, {Rehm}, {Rehm}, {Reid}, {Reis}, {Renk}, {Reoch}, {Ressler}, {Rest},
  {Reynolds}, {Richon}, {Richon}, {Ridgaway}, {Riedel}, {Rieke}, {Rieke}, {Rifelli}, {Rigby}, {Riggs}, {Ringel}, {Ritchie}, {Rix}, {Robberto}, {Robinson}, {Robinson}, {Robinson}, {Rock}, {Rodriguez}, {Rodr{\'\i}guez del Pino}, {Roellig}, {Rohrbach}, {Roman}, {Romelfanger}, {Romo}, {Rosales}, {Rose}, {Roteliuk}, {Roth}, {Rothwell}, {Rouzaud}, {Rowe}, {Rowlands}, {Roy}, {Royer}, {Rui}, {Rumler}, {Rumpl}, {Russ}, {Ryan}, {Ryan}, {Saad}, {Sabata}, {Sabatino}, {Sabbi}, {Sabelhaus}, {Sabia}, {Sahu}, {Saif}, {Salvignol}, {Samara-Ratna}, {Samuelson}, {Sanders}, {Sappington}, {Sargent}, {Sauer}, {Savadkin}, {Sawicki}, {Schappell}, {Scheffer}, {Scheithauer}, {Scherer}, {Schiff}, {Schlawin}, {Schmeitzky}, {Schmitz}, {Schmude}, {Schneider}, {Schreiber}, {Schroeven-Deceuninck}, {Schultz}, {Schwab}, {Schwartz}, {Scoccimarro}, {Scott}, {Scott}, {Seaton}, {Seely}, {Seery}, {Seidleck}, {Sembach}, {Shanahan}, {Shaughnessy}, {Shaw}, {Shay}, {Sheehan}, {Sheth}, {Shih}, {Shivaei}, {Siegel}, {Sienkiewicz}, {Simmons}, {Simon},
  {Sirianni}, {Sivaramakrishnan}, {Slade}, {Sloan}, {Slocum}, {Slowinski}, {Smith}, {Smith}, {Smith}, {Smith}, {Smith}, {Smith}, {Smolik}, {Soderblom}, {Sohn}, {Sokol}, {Sonneborn}, {Sontag}, {Sooy}, {Soummer}, {Southwood}, {Spain}, {Sparmo}, {Speer}, {Spencer}, {Sprofera}, {Stallcup}, {Stanley}, {Stansberry}, {Stark}, {Starr}, {Stassi}, {Steck}, {Steeley}, {Stephens}, {Stephenson}, {Stewart}, {Stiavelli}, {}, {Strada}, {Straughn}, {Streetman}, {Strickland}, {Strobele}, {Stuhlinger}, {Stys}, {Such}, {Sukhatme}, {Sullivan}, {Sullivan}, {Sumner}, {Sun}, {Sunnquist}, {Swade}, {Swam}, {Swenton}, {Swoish}, {Tam Litten}, {Tamas}, {Tao}, {Taylor}, {Taylor}, {te Plate}, {Van Tea}, {Teague}, {Telfer}, {Temim}, {Texter}, {Thatte}, {Thompson}, {Thompson}, {Thomson}, {Thronson}, {Tierney}, {Tikkanen}, {Tinnin}, {Tippet}, {Todd}, {Tran}, {Trauger}, {Trejo}, {Vinh Truong}, {Tsukamoto}, {Tufail}, {Tumlinson}, {Tustain}, {Tyra}, {Ubeda}, {Underwood}, {Uzzo}, {Vaclavik}, {Valenduc}, {Valenti}, {Van Campen}, {van de Wetering},
  {Van Der Marel}, {van Haarlem}, {Vandenbussche}, {van Dishoeck}, {Vanterpool}, {Vernoy}, {Vila Costas}, {Volk}, {Voorzaat}, {Voyton}, {Vydra}, {Waddy}, {Waelkens}, {Wahlgren}, {Walker}, {Wander}, {Warfield}, {Warner}, {Wasiak}, {Wasiak}, {Wehner}, {Weiler}, {Weilert}, {Weiss}, {Wells}, {Welty}, {Wheate}, {Wheeler}, {White}, {Whitehouse}, {Whiteleather}, {Whitman}, {Williams}, {Willmer}, {Willott}, {Willoughby}, {Wilson}, {Wilson}, {Wilson}, {Windhorst}, {Wislowski}, {Wolfe}, {Wolfe}, {Wolff}, {Wondel}, {Woo}, {Woods}, {Worden}, {Workman}, {Wright}, {Wu}, {Wu}, {Wun}, {Wymer}, {Yadetie}, {Yan}, {Yang}, {Yates}, {Yeager}, {Yerger}, {Young}, {Young}, {Yu}, {Yu}, {Zak}, {Zeidler}, {Zepp}, {Zhou}, {Zincke}, {Zonak}, \& {Zondag}}]{2023Gardner}
{Gardner}, J.~P., {Mather}, J.~C., {Abbott}, R., {et~al.} 2023, \pasp, 135, 068001, \dodoi{10.1088/1538-3873/acd1b5}

\bibitem[{{Greenbaum} {et~al.}(2015){Greenbaum}, {Pueyo}, {Sivaramakrishnan}, \& {Lacour}}]{greebaum2015}
{Greenbaum}, A.~Z., {Pueyo}, L., {Sivaramakrishnan}, A., \& {Lacour}, S. 2015, \apj, 798, 68, \dodoi{10.1088/0004-637X/798/2/68}

\bibitem[{{Haniff} {et~al.}(1987){Haniff}, {Mackay}, {Titterington}, {Sivia}, \& {Baldwin}}]{1987Haniff}
{Haniff}, C.~A., {Mackay}, C.~D., {Titterington}, D.~J., {Sivia}, D., \& {Baldwin}, J.~E. 1987, \nat, 328, 694, \dodoi{10.1038/328694a0}

\bibitem[{{Hinkley} {et~al.}(2015){Hinkley}, {Kraus}, {Ireland}, {Cheetham}, {Carpenter}, {Tuthill}, {Lacour}, {Evans}, \& {Haubois}}]{2015Hinkley}
{Hinkley}, S., {Kraus}, A.~L., {Ireland}, M.~J., {et~al.} 2015, \apjl, 806, L9, \dodoi{10.1088/2041-8205/806/1/L9}

\bibitem[{{Hinkley} {et~al.}(2022){Hinkley}, {Carter}, {Ray}, {Skemer}, {Biller}, {Choquet}, {Millar-Blanchaer}, {Sallum}, {Miles}, {Whiteford}, {Patapis}, {Perrin}, {Pueyo}, {Schneider}, {Stapelfeldt}, {Wang}, {Ward-Duong}, {Bowler}, {Boccaletti}, {H. Girard}, {Hines}, {Kalas}, {Kammerer}, {Kervella}, {Leisenring}, {Pantin}, {Zhou}, {Meyer}, {Liu}, {Bonnefoy}, {Currie}, {McElwain}, {Metchev}, {Wyatt}, {Absil}, {Adams}, {Barman}, {Baraffe}, {Bonavita}, {Booth}, {Bryan}, {Chauvin}, {Chen}, {Danielski}, {De Furio}, {Factor}, {Fitzgerald}, {Fortney}, {Grady}, {Greenbaum}, {Henning}, {Hoch}, {Janson}, {Kennedy}, {Kenworthy}, {Kraus}, {Kuzuhara}, {Lagage}, {Lagrange}, {Launhardt}, {Lazzoni}, {Lloyd}, {Marino}, {Marley}, {Martinez}, {Marois}, {Matthews}, {Matthews}, {Mawet}, {Mazoyer}, {Phillips}, {Petrus}, {Quanz}, {Quirrenbach}, {Rameau}, {Rebollido}, {Rickman}, {Samland}, {Sargent}, {Schlieder}, {Sivaramakrishnan}, {Stone}, {Tamura}, {Tremblin}, {Uyama}, {Vasist}, {Vigan}, {Wagner}, \& {Ygouf}}]{2022Hinkley}
{Hinkley}, S., {Carter}, A.~L., {Ray}, S., {et~al.} 2022, \pasp, 134, 095003, \dodoi{10.1088/1538-3873/ac77bd}

\bibitem[{{Hinkley} {et~al.}(2023){Hinkley}, {Lacour}, {Marleau}, {Lagrange}, {Wang}, {Kammerer}, {Cumming}, {Nowak}, {Rodet}, {Stolker}, {Balmer}, {Ray}, {Bonnefoy}, {Molli{\`e}re}, {Lazzoni}, {Kennedy}, {Mordasini}, {Abuter}, {Aigrain}, {Amorim}, {Asensio-Torres}, {Babusiaux}, {Benisty}, {Berger}, {Beust}, {Blunt}, {Boccaletti}, {Bohn}, {Bonnet}, {Bourdarot}, {Brandner}, {Cantalloube}, {Caselli}, {Charnay}, {Chauvin}, {Chomez}, {Choquet}, {Christiaens}, {Cl{\'e}net}, {Coud{\'e} du Foresto}, {Cridland}, {Delorme}, {Dembet}, {Drescher}, {Duvert}, {Eckart}, {Eisenhauer}, {Feuchtgruber}, {Galland}, {Garcia}, {Garcia Lopez}, {Gardner}, {Gendron}, {Genzel}, {Gillessen}, {Girard}, {Grandjean}, {Haubois}, {Hei{\ss}el}, {Henning}, {Hippler}, {Horrobin}, {Houll{\'e}}, {Hubert}, {Jocou}, {Keppler}, {Kervella}, {Kreidberg}, {Lapeyr{\`e}re}, {Le Bouquin}, {L{\'e}na}, {Lutz}, {Maire}, {Mang}, {M{\'e}rand}, {Meunier}, {Monnier}, {Mouillet}, {Nasedkin}, {Ott}, {Otten}, {Paladini}, {Paumard}, {Perraut}, {Perrin},
  {Philipot}, {Pfuhl}, {Pourr{\'e}}, {Pueyo}, {Rameau}, {Rickman}, {Rubini}, {Rustamkulov}, {Samland}, {Shangguan}, {Shimizu}, {Sing}, {Straubmeier}, {Sturm}, {Tacconi}, {van Dishoeck}, {Vigan}, {Vincent}, {Ward-Duong}, {Widmann}, {Wieprecht}, {Wiezorrek}, {Woillez}, {Yazici}, {Young}, \& {Zicher}}]{2023Hinkley}
{Hinkley}, S., {Lacour}, S., {Marleau}, G.~D., {et~al.} 2023, \aap, 671, L5, \dodoi{10.1051/0004-6361/202244727}

\bibitem[{{Hirata} \& {Choi}(2020)}]{2020Hirata}
{Hirata}, C.~M., \& {Choi}, A. 2020, \pasp, 132, 014501, \dodoi{10.1088/1538-3873/ab44f7}

\bibitem[{{Ireland}(2013)}]{2013Ireland}
{Ireland}, M.~J. 2013, \mnras, 433, 1718, \dodoi{10.1093/mnras/stt859}

\bibitem[{{Kammerer} {et~al.}(2019){Kammerer}, {Ireland}, {Martinache}, \& {Girard}}]{2019MNRAS.486..639K}
{Kammerer}, J., {Ireland}, M.~J., {Martinache}, F., \& {Girard}, J.~H. 2019, \mnras, 486, 639, \dodoi{10.1093/mnras/stz882}

\bibitem[{{Kammerer} {et~al.}(2023){Kammerer}, {Cooper}, {Vandal}, {Thatte}, {Martinache}, {Sivaramakrishnan}, {Chaushev}, {Stolker}, {Lloyd}, {Albert}, {Doyon}, {Sallum}, {Perrin}, {Pueyo}, {M{\'e}rand}, {Gallenne}, {Greenbaum}, {Sanchez-Bermudez}, {Blakely}, {Johnstone}, {Volk}, {Martel}, {Goudfrooij}, {Meyer}, {Willott}, {De Furio}, {Dang}, {Radica}, \& {Noirot}}]{2023Kammerer}
{Kammerer}, J., {Cooper}, R.~A., {Vandal}, T., {et~al.} 2023, \pasp, 135, 014502, \dodoi{10.1088/1538-3873/ac9a74}

\bibitem[{{Konopacky} {et~al.}(2013){Konopacky}, {Barman}, {Macintosh}, \& {Marois}}]{2013Konopacky}
{Konopacky}, Q.~M., {Barman}, T.~S., {Macintosh}, B.~A., \& {Marois}, C. 2013, Science, 339, 1398, \dodoi{10.1126/science.1232003}

\bibitem[{{Kraus} \& {Ireland}(2012)}]{2012Kraus}
{Kraus}, A.~L., \& {Ireland}, M.~J. 2012, \apj, 745, 5, \dodoi{10.1088/0004-637X/745/1/5}

\bibitem[{{Kraus} {et~al.}(2022){Kraus}, {Mortimer}, {Chhabra}, {Lu}, {Codron}, {Gardner}, {Anugu}, {Monnier}, {Le Bouquin}, {Ireland}, {Martinache}, {Defr{\`e}re}, \& {Martinod}}]{2022Kraus}
{Kraus}, S., {Mortimer}, D., {Chhabra}, S., {et~al.} 2022, in Society of Photo-Optical Instrumentation Engineers (SPIE) Conference Series, Vol. 12183, Optical and Infrared Interferometry and Imaging VIII, ed. A.~{M{\'e}rand}, S.~{Sallum}, \& J.~{Sanchez-Bermudez}, 121831S, \dodoi{10.1117/12.2627973}

\bibitem[{{Lacour} {et~al.}(2019){Lacour}, {Dembet}, {Abuter}, {F{\'e}dou}, {Perrin}, {Choquet}, {Pfuhl}, {Eisenhauer}, {Woillez}, {Cassaing}, {Wieprecht}, {Ott}, {Wiezorrek}, {Tristram}, {Wolff}, {Ram{\'\i}rez}, {Haubois}, {Perraut}, {Straubmeier}, {Brandner}, \& {Amorim}}]{2019Lacour}
{Lacour}, S., {Dembet}, R., {Abuter}, R., {et~al.} 2019, \aap, 624, A99, \dodoi{10.1051/0004-6361/201834981}

\bibitem[{{Lau} {et~al.}(2023){Lau}, {Hankins}, {Sanchez-Bermudez}, {Thatte}, {Soulain}, {Cooper}, {Sivaramakrishnan}, {Corcoran}, {Greenbaum}, {Gull}, {Han}, {Jones}, {Madura}, {Moffat}, {Morris}, {Onaka}, {Russell}, {Richardson}, {Smith}, {Tuthill}, {Volk}, {Weigelt}, \& {Williams}}]{2023Lau}
{Lau}, R.~M., {Hankins}, M.~J., {Sanchez-Bermudez}, J., {et~al.} 2023, arXiv e-prints, arXiv:2311.15948, \dodoi{10.48550/arXiv.2311.15948}

\bibitem[{{Marois} {et~al.}(2010){Marois}, {Zuckerman}, {Konopacky}, {Macintosh}, \& {Barman}}]{2010Marois}
{Marois}, C., {Zuckerman}, B., {Konopacky}, Q.~M., {Macintosh}, B., \& {Barman}, T. 2010, \nat, 468, 1080, \dodoi{10.1038/nature09684}

\bibitem[{{Miles} {et~al.}(2023){Miles}, {Biller}, {Patapis}, {Worthen}, {Rickman}, {Hoch}, {Skemer}, {Perrin}, {Whiteford}, {Chen}, {Sargent}, {Mukherjee}, {Morley}, {Moran}, {Bonnefoy}, {Petrus}, {Carter}, {Choquet}, {Hinkley}, {Ward-Duong}, {Leisenring}, {Millar-Blanchaer}, {Pueyo}, {Ray}, {Sallum}, {Stapelfeldt}, {Stone}, {Wang}, {Absil}, {Balmer}, {Boccaletti}, {Bonavita}, {Booth}, {Bowler}, {Chauvin}, {Christiaens}, {Currie}, {Danielski}, {Fortney}, {Girard}, {Grady}, {Greenbaum}, {Henning}, {Hines}, {Janson}, {Kalas}, {Kammerer}, {Kennedy}, {Kenworthy}, {Kervella}, {Lagage}, {Lew}, {Liu}, {Macintosh}, {Marino}, {Marley}, {Marois}, {Matthews}, {Matthews}, {Mawet}, {McElwain}, {Metchev}, {Meyer}, {Molliere}, {Pantin}, {Quirrenbach}, {Rebollido}, {Ren}, {Schneider}, {Vasist}, {Wyatt}, {Zhou}, {Briesemeister}, {Bryan}, {Calissendorff}, {Cantalloube}, {Cugno}, {De Furio}, {Dupuy}, {Factor}, {Faherty}, {Fitzgerald}, {Franson}, {Gonzales}, {Hood}, {Howe}, {Kraus}, {Kuzuhara}, {Lagrange}, {Lawson}, {Lazzoni},
  {Liu}, {Llop-Sayson}, {Lloyd}, {Martinez}, {Mazoyer}, {Quanz}, {Redai}, {Samland}, {Schlieder}, {Tamura}, {Tan}, {Uyama}, {Vigan}, {Vos}, {Wagner}, {Wolff}, {Ygouf}, {Zhang}, {Zhang}, \& {Zhang}}]{2023Miles}
{Miles}, B.~E., {Biller}, B.~A., {Patapis}, P., {et~al.} 2023, \apjl, 946, L6, \dodoi{10.3847/2041-8213/acb04a}

\bibitem[{{Monnier} {et~al.}(2007){Monnier}, {Tuthill}, {Danchi}, {Murphy}, \& {Harries}}]{2007Monnier}
{Monnier}, J.~D., {Tuthill}, P.~G., {Danchi}, W.~C., {Murphy}, N., \& {Harries}, T.~J. 2007, \apj, 655, 1033, \dodoi{10.1086/509873}

\bibitem[{{Nowak} {et~al.}(2020){Nowak}, {Lacour}, {Lagrange}, {Rubini}, {Wang}, {Stolker}, {Abuter}, {Amorim}, {Asensio-Torres}, {Baub{\"o}ck}, {Benisty}, {Berger}, {Beust}, {Blunt}, {Boccaletti}, {Bonnefoy}, {Bonnet}, {Brandner}, {Cantalloube}, {Charnay}, {Choquet}, {Christiaens}, {Cl{\'e}net}, {Coud{\'e} Du Foresto}, {Cridland}, {de Zeeuw}, {Dembet}, {Dexter}, {Drescher}, {Duvert}, {Eckart}, {Eisenhauer}, {Gao}, {Garcia}, {Garcia Lopez}, {Gardner}, {Gendron}, {Genzel}, {Gillessen}, {Girard}, {Grandjean}, {Haubois}, {Hei{\ss}el}, {Henning}, {Hinkley}, {Hippler}, {Horrobin}, {Houll{\'e}}, {Hubert}, {Jim{\'e}nez-Rosales}, {Jocou}, {Kammerer}, {Kervella}, {Keppler}, {Kreidberg}, {Kulikauskas}, {Lapeyr{\`e}re}, {Le Bouquin}, {L{\'e}na}, {M{\'e}rand}, {Maire}, {Molli{\`e}re}, {Monnier}, {Mouillet}, {M{\"u}ller}, {Nasedkin}, {Ott}, {Otten}, {Paumard}, {Paladini}, {Perraut}, {Perrin}, {Pueyo}, {Pfuhl}, {Rameau}, {Rodet}, {Rodr{\'\i}guez-Coira}, {Rousset}, {Scheithauer}, {Shangguan}, {Stadler}, {Straub},
  {Straubmeier}, {Sturm}, {Tacconi}, {van Dishoeck}, {Vigan}, {Vincent}, {von Fellenberg}, {Ward-Duong}, {Widmann}, {Wieprecht}, {Wiezorrek}, {Woillez}, \& {Gravity Collaboration}}]{2020Nowak}
{Nowak}, M., {Lacour}, S., {Lagrange}, A.~M., {et~al.} 2020, \aap, 642, L2, \dodoi{10.1051/0004-6361/202039039}

\bibitem[{{Phillips} {et~al.}(2020){Phillips}, {Tremblin}, {Baraffe}, {Chabrier}, {Allard}, {Spiegelman}, {Goyal}, {Drummond}, \& {H{\'e}brard}}]{PhillipsPaper}
{Phillips}, M.~W., {Tremblin}, P., {Baraffe}, I., {et~al.} 2020, \aap, 637, A38, \dodoi{10.1051/0004-6361/201937381}

\bibitem[{{Ray} {et~al.}(2023){Ray}, {Hinkley}, {Sallum}, {Bonavita}, {Squicciarini}, {Carter}, \& {Lazzoni}}]{2022Ray}
{Ray}, S., {Hinkley}, S., {Sallum}, S., {et~al.} 2023, \mnras, 519, 2718, \dodoi{10.1093/mnras/stac3425}

\bibitem[{{Readhead} {et~al.}(1988){Readhead}, {Nakajima}, {Pearson}, {Neugebauer}, {Oke}, \& {Sargent}}]{1988Readhead}
{Readhead}, A.~C.~S., {Nakajima}, T.~S., {Pearson}, T.~J., {et~al.} 1988, \aj, 95, 1278, \dodoi{10.1086/114724}

\bibitem[{{Rigby} {et~al.}(2022){Rigby}, {Perrin}, {McElwain}, {Kimble}, {Friedman}, {Lallo}, {Doyon}, {Feinberg}, {Ferruit}, {Glasse}, {Rieke}, {Rieke}, {Wright}, {Willott}, {Colon}, {Milam}, {Neff}, {Stark}, {Valenti}, {Abell}, {Abney}, {Abul-Huda}, {Acton}, {Adams}, {Adler}, {Aguilar}, {Ahmed}, {Albert}, {Alberts}, {Aldridge}, {Allen}, {Altenburg}, {Alves de Oliveira}, {Anderson}, {Anderson}, {Anderson}, {Argyriou}, {Armstrong}, {Arribas}, {Artigau}, {Arvai}, {Atkinson}, {Bacon}, {Bair}, {Banks}, {Barrientes}, {Barringer}, {Bartosik}, {Bast}, {Baudoz}, {Beatty}, {Bechtold}, {Beck}, {Bergeron}, {Bergkoetter}, {Bhatawdekar}, {Birkmann}, {Blazek}, {Blome}, {Boccaletti}, {Boeker}, {Boia}, {Bonaventura}, {Bond}, {Bosley}, {Boucarut}, {Bourque}, {Bouwman}, {Bower}, {Bowers}, {Boyer}, {Brady}, {Braun}, {Breda}, {Bresnahan}, {Bright}, {Britt}, {Bromenschenkel}, {Brooks}, {Brooks}, {Brown}, {Brown}, {Brown}, {Bunker}, {Burger}, {Bushouse}, {Cale}, {Cameron}, {Cameron}, {Canipe}, {Caplinger}, {Caputo}, {Carey},
  {Carniani}, {Carrasquilla}, {Carruthers}, {Case}, {Chance}, {Chapman}, {Charlot}, {Charlow}, {Chayer}, {Chen}, {Cherinka}, {Chichester}, {Chilton}, {Chonis}, {Clark}, {Clark}, {Coe}, {Coleman}, {Comber}, {Comeau}, {Connolly}, {Cooper}, {Cooper}, {Coppock}, {Correnti}, {Cossou}, {Coulais}, {Coyle}, {Cracraft}, {Curti}, {Cuturic}, {Davis}, {Davis}, {Dean}, {DeLisa}, {deMeester}, {Dencheva}, {Dencheva}, {DePasquale}, {Deschenes}, {Hunor Detre}, {Diaz}, {Dicken}, {DiFelice}, {Dillman}, {Dixon}, {Doggett}, {Donaldson}, {Douglas}, {DuPrie}, {Dupuis}, {Durning}, {Easmin}, {Eck}, {Edeani}, {Egami}, {Ehrenwinkler}, {Eisenhamer}, {Eisenhower}, {Elie}, {Elliott}, {Elliott}, {Ellis}, {Engesser}, {Espinoza}, {Etienne}, {Etxaluze}, {Falini}, {Feeney}, {Ferry}, {Filippazzo}, {Fincham}, {Fix}, {Flagey}, {Florian}, {Flynn}, {Fontanella}, {Ford}, {Forshay}, {Fox}, {Franz}, {Fu}, {Fullerton}, {Galkin}, {Galyer}, {Garcia Marin}, {Gardner}, {Gardner}, {Garland}, {Gasman}, {Gaspar}, {Gaudreau}, {Gauthier}, {Geers}, {Geithner},
  {Gennaro}, {Giardino}, {Girard}, {Giuliano}, {Glassmire}, {Glauser}, {Glazer}, {Godfrey}, {Golimowski}, {Gollnitz}, {Gong}, {Gonzaga}, {Gordon}, {Gordon}, {Goudfrooij}, {Greene}, {Greenhouse}, {Grimaldi}, {Groebner}, {Grundy}, {Guillard}, {Gutman}, {Ha}, {Haderlein}, {Hagedorn}, {Hainline}, {Haley}, {Hami}, {Hamilton}, {Hammel}, {Hansen}, {Harkins}, {Harr}, {Hart}, {Hart}, {Hartig}, {Hashimoto}, {Haskins}, {Hathaway}, {Havey}, {Hayden}, {Hecht}, {Heller-Boyer}, {Henry}, {Hermann}, {Hernandez}, {Hesman}, {Hicks}, {Hilbert}, {Hines}, {Hoffman}, {Holfeltz}, {Holler}, {Hoppa}, {Hott}, {Howard}, {Hunter}, {Hunter}, {Hurst}, {Husemann}, {Hustak}, {Ilinca Ignat}, {Irish}, {Jackson}, {Jahromi}, {Jakobsen}, {James}, {James}, {Januszewski}, {Jenkins}, {Jirdeh}, {Johnson}, {Johnson}, {Jones}, {Jones}, {Jones}, {Jones}, {Jordan}, {Jordan}, {Jurczyk}, {Jurling}, {Kaleida}, {Kalmanson}, {Kammerer}, {Kang}, {Kao}, {Karakla}, {Kavanagh}, {Kelly}, {Kendrew}, {Kennedy}, {Kenny}, {Keski-kuha}, {Keyes}, {Kidwell}, {Kinzel},
  {Kirk}, {Kirkpatrick}, {Kirshenblat}, {Klaassen}, {Knapp}, {Knight}, {Knollenberg}, {Koehler}, {Koekemoer}, {Kovacs}, {Kulp}, {Kumari}, {Kyprianou}, {La Massa}, {Labador}, {Labiano Ortega}, {Lagage}, {Lajoie}, {Lallo}, {Lam}, {Lamb}, {Lambros}, {Lampenfield}, {Langston}, {Larson}, {Law}, {Lawrence}, {Lee}, {Leisenring}, {Lepo}, {Leveille}, {Levenson}, {Levine}, {Levy}, {Lewis}, {Lewis}, {Libralato}, {Lightsey}, {Link}, {Liu}, {Lo}, {Lockwood}, {Logue}, {Long}, {Long}, {Loomis}, {Lopez-Caniego}, {Alvarez}, {Love-Pruitt}, {Lucy}, {Luetzgendorf}, {Maghami}, {Maiolino}, {Major}, {Malla}, {Malumuth}, {Manjavacas}, {Mannfolk}, {Marrione}, {Marston}, {Martel}, {Maschmann}, {Masci}, {Masciarelli}, {Maszkiewicz}, {Mather}, {McKenzie}, {McLean}, {McMaster}, {Melbourne}, {Mel{\'e}ndez}, {Menzel}, {Merz}, {Meyett}, {Meza}, {Miskey}, {Misselt}, {Moller}, {Morrison}, {Morse}, {Moseley}, {Mosier}, {Mountain}, {Mueckay}, {Mueller}, {Mullally}, {Murphy}, {Murray}, {Murray}, {Muzerolle}, {Mycroft}, {Myers}, {Myrick},
  {Nanavati}, {Nance}, {Nayak}, {Naylor}, {Nelan}, {Nickson}, {Nielson}, {Nieto-Santisteban}, {Nikolov}, {Noriega-Crespo}, {O'Shaughnessy}, {O'Sullivan}, {Ochs}, {Ogle}, {Oleszczuk}, {Olmsted}, {Osborne}, {Ottens}, {Owens}, {Pacifici}, {Pagan}, {Page}, {Parrish}, {Patapis}, {Pauly}, {Pavlovsky}, {Pedder}, {Peek}, {Pena-Guerrero}, {Pennanen}, {Perez}, {Perna}, {Perriello}, {Phillips}, {Pietraszkiewicz}, {Pinaud}, {Pirzkal}, {Pitman}, {Piwowar}, {Platais}, {Player}, {Plesha}, {Pollizi}, {Polster}, {Pontoppidan}, {Porterfield}, {Proffitt}, {Pueyo}, {Pulliam}, {Quirt}, {Quispe Neira}, {Ramos Alarcon}, {Ramsay}, {Rapp}, {Rapp}, {Rauscher}, {Ravindranath}, {Rawle}, {Regan}, {Reichard}, {Reis}, {Ressler}, {Rest}, {Reynolds}, {Rhue}, {Richon}, {Rickman}, {Ridgaway}, {Ritchie}, {Rix}, {Robberto}, {Robinson}, {Robinson}, {Robinson}, {Rock}, {Rodriguez}, {Rodriguez Del Pino}, {Roellig}, {Rohrbach}, {Roman}, {Romelfanger}, {Rose}, {Roteliuk}, {Roth}, {Rothwell}, {Rowlands}, {Roy}, {Royer}, {Royle}, {Rui}, {Rumler},
  {Runnels}, {Russ}, {Rustamkulov}, {Ryden}, {Ryer}, {Sabata}, {Sabatke}, {Sabbi}, {Samuelson}, {Sappington}, {Sargent}, {Sauer}, {Scheithauer}, {Schlawin}, {Schlitz}, {Schmitz}, {Schneider}, {Schreiber}, {Schulze}, {Schwab}, {Scott}, {Sembach}, {Shaughnessy}, {Shaw}, {Shawger}, {Shay}, {Sheehan}, {Shen}, {Sherman}, {Shiao}, {Shih}, {Shivaei}, {Sienkiewicz}, {Sing}, {Sirianni}, {Sivaramakrishnan}, {Skipper}, {Sloan}, {Slocum}, {Slowinski}, {Smith}, {Smith}, {Smith}, {Smith}, {Snyder}, {Soh}, {Sohn}, {Soto}, {Spencer}, {Stallcup}, {Stansberry}, {Starr}, {Starr}, {Stewart}, {Stiavelli}, {Straughn}, {Strickland}, {Stys}, {Summers}, {Sun}, {Sunnquist}, {Swade}, {Swam}, {Swaters}, {Swoish}, {Taylor}, {Taylor}, {Te Plate}, {Tea}, {Teague}, {Telfer}, {Temim}, {Thatte}, {Thompson}, {Thompson}, {Thomson}, {Tikkanen}, {Tippet}, {Todd}, {Toolan}, {Tran}, {Trejo}, {Truong}, {Tsukamoto}, {Tustain}, {Tyra}, {Ubeda}, {Underwood}, {Uzzo}, {Van Campen}, {Vandal}, {Vandenbussche}, {Vila}, {Volk}, {Wahlgren}, {Waldman},
  {Walker}, {Wander}, {Warfield}, {Warner}, {Wasiak}, {Watkins}, {Weilert}, {Weiser}, {Weiss}, {Weissman}, {Welty}, {West}, {Wheate}, {Wheatley}, {Wheeler}, {White}, {Whiteaker}, {Whitehouse}, {Whiteleather}, {Whitman}, {Williams}, {Willmer}, {Willoughby}, {Wilson}, {Wirth}, {Wislowski}, {Wolf}, {Wolfe}, {Wolff}, {Workman}, {Wright}, {Wu}, {Wu}, {Wymer}, {Yates}, {Yates}, {Yeager}, {Yerger}, {Yoon}, {Young}, {Yu}, {Zak}, {Zeidler}, {Zhou}, {Zielinski}, {Zincke}, \& {Zonak}}]{2022JWSTCommissioning}
{Rigby}, J., {Perrin}, M., {McElwain}, M., {et~al.} 2022, arXiv e-prints, arXiv:2207.05632.
\newblock \doarXiv{2207.05632}

\bibitem[{{Sallum} {et~al.}(2022){Sallum}, {Ray}, \& {Hinkley}}]{StephSPIE}
{Sallum}, S., {Ray}, S., \& {Hinkley}, S. 2022, in Society of Photo-Optical Instrumentation Engineers (SPIE) Conference Series, Vol. 12183, Optical and Infrared Interferometry and Imaging VIII, ed. A.~{M{\'e}rand}, S.~{Sallum}, \& J.~{Sanchez-Bermudez}, 121832M, \dodoi{10.1117/12.2630401}

\bibitem[{{Sallum} {et~al.}(2019){Sallum}, {Bailey}, {Bernstein}, {Boss}, {Bowler}, {Close}, {Currie}, {Dong}, {Espaillat}, {Fitzgerald}, {Follette}, {Fortney}, {Hasegawa}, {Jang-Condell}, {Jovanovic}, {Kane}, {Konopacky}, {Liu}, {Lozi}, {Males}, {Mawet}, {Mazin}, {Millar-Blanchaer}, {Murray-Clay}, {Ruane}, {Skemer}, {Tamura}, {Vasisht}, {Wang}, \& {Wang}}]{2019Sallum}
{Sallum}, S., {Bailey}, V., {Bernstein}, R.~A., {et~al.} 2019, \baas, 51, 527.
\newblock \doarXiv{1903.05319}

\bibitem[{{Sallum} {et~al.}(2024){Sallum}, {Ray}, {Kammerer}, {Sivaramakrishnan}, {Cooper}, {Greebaum}, {Thatte}, {De Furio}, {Factor}, {Meyer}, {Stone}, {Carter}, {Biller}, {Hinkley}, {Skemer}, {Su{\'a}rez}, {Leisenring}, {Perrin}, {Kraus}, {Absil}, {Balmer}, {Betti}, {Boccaletti}, {Bonavita}, {Bonnefoy}, {Booth}, {Bowler}, {Briesemeister}, {Bryan}, {Calissendorff}, {Cantalloube}, {Chauvin}, {Chen}, {Choquet}, {Christiaens}, {Cugno}, {Currie}, {Danielski}, {Dupuy}, {Faherty}, {Fitzgerald}, {Fortney}, {Franson}, {Girard}, {Grady}, {Gonzales}, {Henning}, {Hines}, {Hoch}, {Hood}, {Howe}, {Janson}, {Kalas}, {Kennedy}, {Kenworthy}, {Kervella}, {Kitzmann}, {Kuzuhara}, {Lagrange}, {Lagage}, {Lawson}, {Lazzoni}, {Lew}, {Liu}, {Liu}, {Llop-Sayson}, {Lloyd}, {Lueber}, {Macintosh}, {Manjavacas}, {Marino}, {Marley}, {Marois}, {Martinez}, {Matthews}, {Matthews}, {Mawet}, {Mazoyer}, {McElwain}, {Metchev}, {Miles}, {Millar-Blanchaer}, {Molliere}, {Moran}, {Morley}, {Mukherjee}, {Palma-Bifani}, {Pantin}, {Patapis},
  {Petrus}, {Pueyo}, {Quanz}, {Quirrenbach}, {Rebollido}, {Redai}, {Ren}, {Rickman}, {Samland}, {Sargent}, {Schlieder}, {Schneider}, {Stapelfeldt}, {Sutlieff}, {Tamura}, {Tan}, {Theissen}, {Uyama}, {Vigan}, {Vasist}, {Vos}, {Wagner}, {Wang}, {Ward-Duong}, {Whiteford}, {Wolff}, {Worthen}, {Wyatt}, {Ygouf}, {Zhang}, {Zhang}, {Zhang}, {Zhou}, \& {Zurlo}}]{sallum_subm}
{Sallum}, S., {Ray}, S., {Kammerer}, J., {et~al.} 2024, \apjl, 963, L2, \dodoi{10.3847/2041-8213/ad21fb}

\bibitem[{{Saumon} \& {Marley}(2008)}]{2008Saumon}
{Saumon}, D., \& {Marley}, M.~S. 2008, \apj, 689, 1327, \dodoi{10.1086/592734}

\bibitem[{{Sivaramakrishnan} {et~al.}(2009){Sivaramakrishnan}, {Tuthill}, {Ireland}, {Lloyd}, {Martinache}, {Soummer}, {Makidon}, {Doyon}, {Beaulieu}, \& {Beichman}}]{SivaramkrishnanNRM}
{Sivaramakrishnan}, A., {Tuthill}, P.~G., {Ireland}, M.~J., {et~al.} 2009, in Society of Photo-Optical Instrumentation Engineers (SPIE) Conference Series, Vol. 7440, Techniques and Instrumentation for Detection of Exoplanets IV, ed. S.~B. {Shaklan}, 74400Y, \dodoi{10.1117/12.826633}

\bibitem[{{Sivaramakrishnan} {et~al.}(2012){Sivaramakrishnan}, {Lafreni{\`e}re}, {Ford}, {McKernan}, {Cheetham}, {Greenbaum}, {Tuthill}, {Lloyd}, {Ireland}, {Doyon}, {Beaulieu}, {Martel}, {Koekemoer}, {Martinache}, \& {Teuben}}]{2012Sivaramkrishnan}
{Sivaramakrishnan}, A., {Lafreni{\`e}re}, D., {Ford}, K.~E.~S., {et~al.} 2012, in Society of Photo-Optical Instrumentation Engineers (SPIE) Conference Series, Vol. 8442, Space Telescopes and Instrumentation 2012: Optical, Infrared, and Millimeter Wave, ed. M.~C. {Clampin}, G.~G. {Fazio}, H.~A. {MacEwen}, \& J.~{Oschmann}, Jacobus~M., 84422S, \dodoi{10.1117/12.925565}

\bibitem[{{Sivaramakrishnan} {et~al.}(2023{\natexlab{a}}){Sivaramakrishnan}, {Tuthill}, {Lloyd}, {Greenbaum}, {Thatte}, {Cooper}, {Vandal}, {Kammerer}, {Sanchez-Bermudez}, {Pope}, {Blakely}, {Albert}, {Cook}, {Johnstone}, {Martel}, {Volk}, {Soulain}, {Artigau}, {Lafreni{\`e}re}, {Willott}, {Parmentier}, {Ford}, {McKernan}, {Vila}, {Rowlands}, {Doyon}, {Beaulieu}, {Desdoigts}, {Fullerton}, {De Furio}, {Goudfrooij}, {Holfeltz}, {LaMassa}, {Maszkiewicz}, {Meyer}, {Perrin}, {Pueyo}, {Sahlmann}, {Sohn}, {Teixeira}, \& {Zheng}}]{2023Sivaramkrishnan}
{Sivaramakrishnan}, A., {Tuthill}, P., {Lloyd}, J.~P., {et~al.} 2023{\natexlab{a}}, \pasp, 135, 015003, \dodoi{10.1088/1538-3873/acaebd}

\bibitem[{{Sivaramakrishnan} {et~al.}(2023{\natexlab{b}}){Sivaramakrishnan}, {Tuthill}, {Lloyd}, {Greenbaum}, {Thatte}, {Cooper}, {Vandal}, {Kammerer}, {Sanchez-Bermudez}, {Pope}, {Blakely}, {Albert}, {Cook}, {Johnstone}, {Martel}, {Volk}, {Soulain}, {Artigau}, {Lafreni{\`e}re}, {Willott}, {Parmentier}, {Ford}, {McKernan}, {Vila}, {Rowlands}, {Doyon}, {Beaulieu}, {Desdoigts}, {Fullerton}, {De Furio}, {Goudfrooij}, {Holfeltz}, {LaMassa}, {Maszkiewicz}, {Meyer}, {Perrin}, {Pueyo}, {Sahlmann}, {Sohn}, {Teixeira}, \& {Zheng}}]{2023Sivaramakrishnan}
---. 2023{\natexlab{b}}, \pasp, 135, 015003, \dodoi{10.1088/1538-3873/acaebd}

\bibitem[{{Soulain} {et~al.}(2020){Soulain}, {Sivaramakrishnan}, {Tuthill}, {Thatte}, {Volk}, {Cooper}, {Albert}, {Artigau}, {Cook}, {Doyon}, {Johnstone}, {Lafreni{\`e}re}, \& {Martel}}]{2020SoulainSPIE}
{Soulain}, A., {Sivaramakrishnan}, A., {Tuthill}, P., {et~al.} 2020, in Society of Photo-Optical Instrumentation Engineers (SPIE) Conference Series, Vol. 11446, Society of Photo-Optical Instrumentation Engineers (SPIE) Conference Series, 1144611, \dodoi{10.1117/12.2560804}

\bibitem[{{Spiegel} \& {Burrows}(2012)}]{2012Spiegel}
{Spiegel}, D.~S., \& {Burrows}, A. 2012, \apj, 745, 174, \dodoi{10.1088/0004-637X/745/2/174}

\bibitem[{{Spiegel} {et~al.}(2011){Spiegel}, {Burrows}, \& {Milsom}}]{2011Spiegel}
{Spiegel}, D.~S., {Burrows}, A., \& {Milsom}, J.~A. 2011, \apj, 727, 57, \dodoi{10.1088/0004-637X/727/1/57}

\bibitem[{{Stolker} {et~al.}(2020){Stolker}, {Quanz}, {Todorov}, {K{\"u}hn}, {Molli{\`e}re}, {Meyer}, {Currie}, {Daemgen}, \& {Lavie}}]{2020Stolker}
{Stolker}, T., {Quanz}, S.~P., {Todorov}, K.~O., {et~al.} 2020, \aap, 635, A182, \dodoi{10.1051/0004-6361/201937159}

\bibitem[{{Tokovinin} {et~al.}(2010){Tokovinin}, {Cantarutti}, {Tighe}, {Schurter}, {van der Bliek}, {Martinez}, \& {Mondaca}}]{2010Tokovinin}
{Tokovinin}, A., {Cantarutti}, R., {Tighe}, R., {et~al.} 2010, \pasp, 122, 1483, \dodoi{10.1086/657903}

\bibitem[{{Tuthill} {et~al.}(2001){Tuthill}, {Monnier}, \& {Danchi}}]{2001Tuthill}
{Tuthill}, P.~G., {Monnier}, J.~D., \& {Danchi}, W.~C. 2001, \nat, 409, 1012, \dodoi{10.1038/35059014}

\bibitem[{{Tuthill} {et~al.}(2000){Tuthill}, {Monnier}, {Danchi}, {Wishnow}, \& {Haniff}}]{2000Tuthill}
{Tuthill}, P.~G., {Monnier}, J.~D., {Danchi}, W.~C., {Wishnow}, E.~H., \& {Haniff}, C.~A. 2000, \pasp, 112, 555, \dodoi{10.1086/316550}

\bibitem[{{Vides} {et~al.}(2023){Vides}, {Sallum}, {Eisner}, {Skemer}, \& {Murray-Clay}}]{2023Vides}
{Vides}, C.~L., {Sallum}, S., {Eisner}, J., {Skemer}, A., \& {Murray-Clay}, R. 2023, \apj, 958, 123, \dodoi{10.3847/1538-4357/acfda6}

\bibitem[{{Wagner} {et~al.}(2019){Wagner}, {Apai}, \& {Kratter}}]{Wagner:2019ApJ}
{Wagner}, K., {Apai}, D., \& {Kratter}, K.~M. 2019, ApJ, 877, 46, \dodoi{10.3847/1538-4357/ab1904}

\end{thebibliography}
\bibliographystyle{aasjournal}




{\appendix}
\renewcommand{\thesection}{\Alph{section}}

\medskip

\section{{NISRAPID Readout pattern}}
\label{append:GroupsAndIntegrations}
\begin{figure*}[h!]
    \centering
    \includegraphics[scale=0.9]{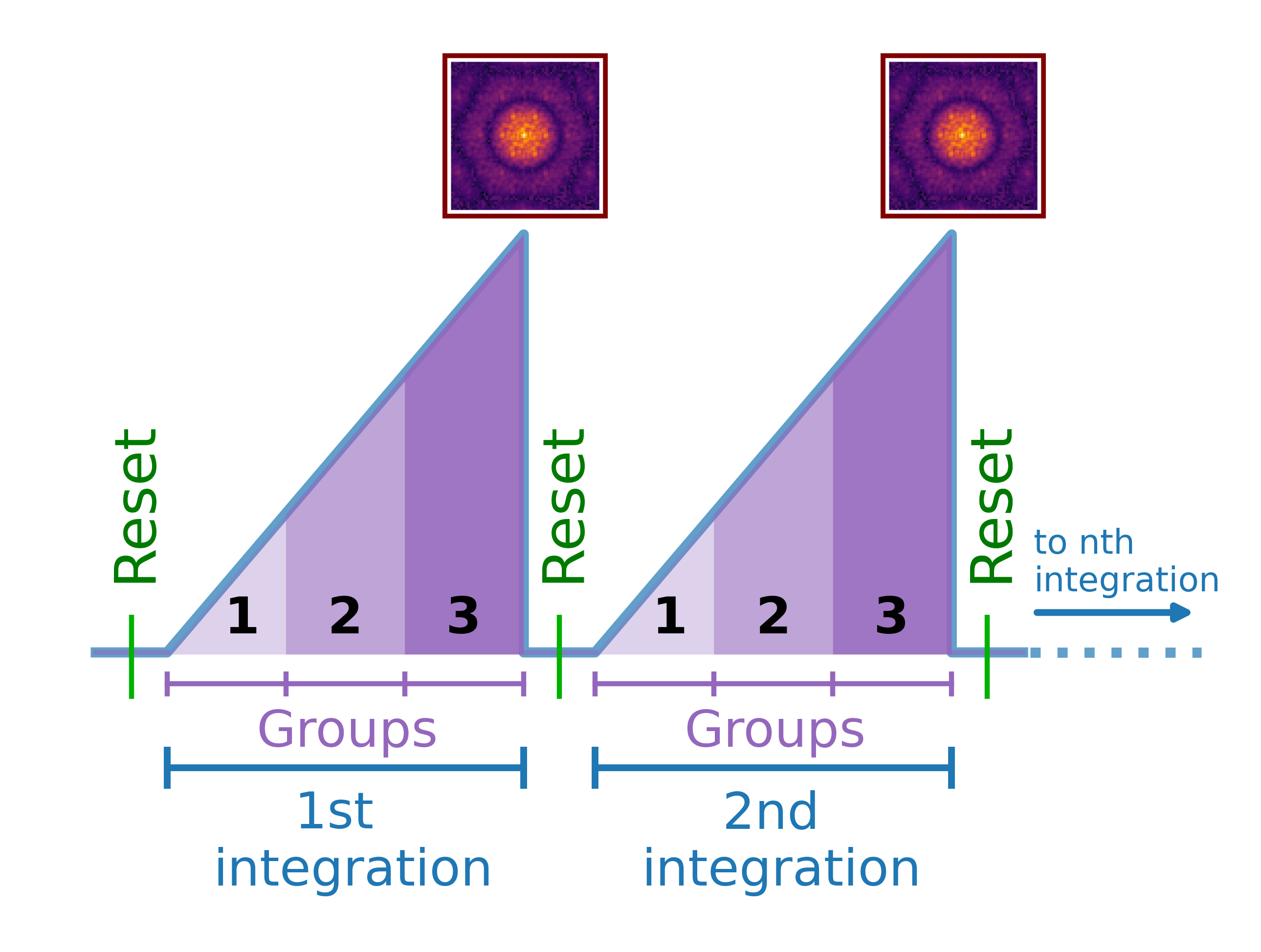}
    \caption{{Groups and integrations in the `NISRAPID' mode which was used by all the observations in this work. Each purple segment represents a `frame' which is the same as a `group' in this mode and is the smallest unit of exposure. An `integration' is comprised of a number of groups (3 in this case) and represents the total number of photons collected in a contiguous sequence. After each integration, the detector resets and executes the next integration to collect photons. The total exposure of an observation is the collection of `$n$' such integrations.}}
    \label{fig:GroupsAndIntegrations}
\end{figure*}

\newpage
\section{{Closure Phases of injected companion signals}}
\label{append:closurephases}
\begin{figure*}[h!]
    \centering
    \includegraphics[scale=0.8]{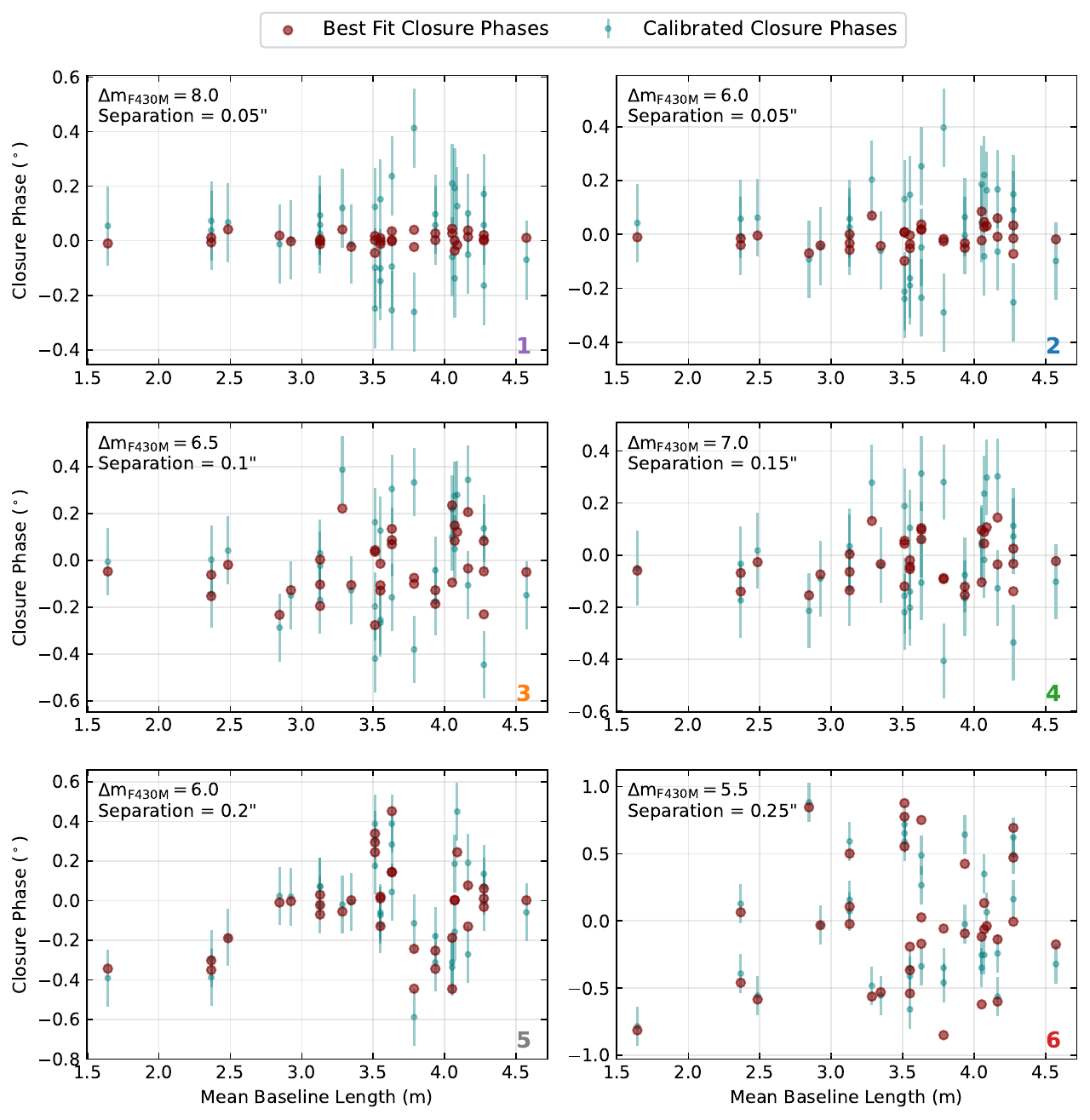}
    \caption{The calibrated closure phases and the corresponding best fits for the injected companion signals. The \textbf{\color{blue}test} numbers are colour coded according to Figure \ref{fig:ContrastsChisqComp} at the bottom right for each plot. The $\Delta m_{F430M}$ and separation values for each injection is also provided on the top left for each plot.}
    \label{fig:InjectedCPs}
\end{figure*}

\newpage

\section{{Squared Visibilities of injected companion signals}}
\label{append:squaredvisibilities}
\begin{figure*}[h!]
    \centering
    \includegraphics[scale=0.8]{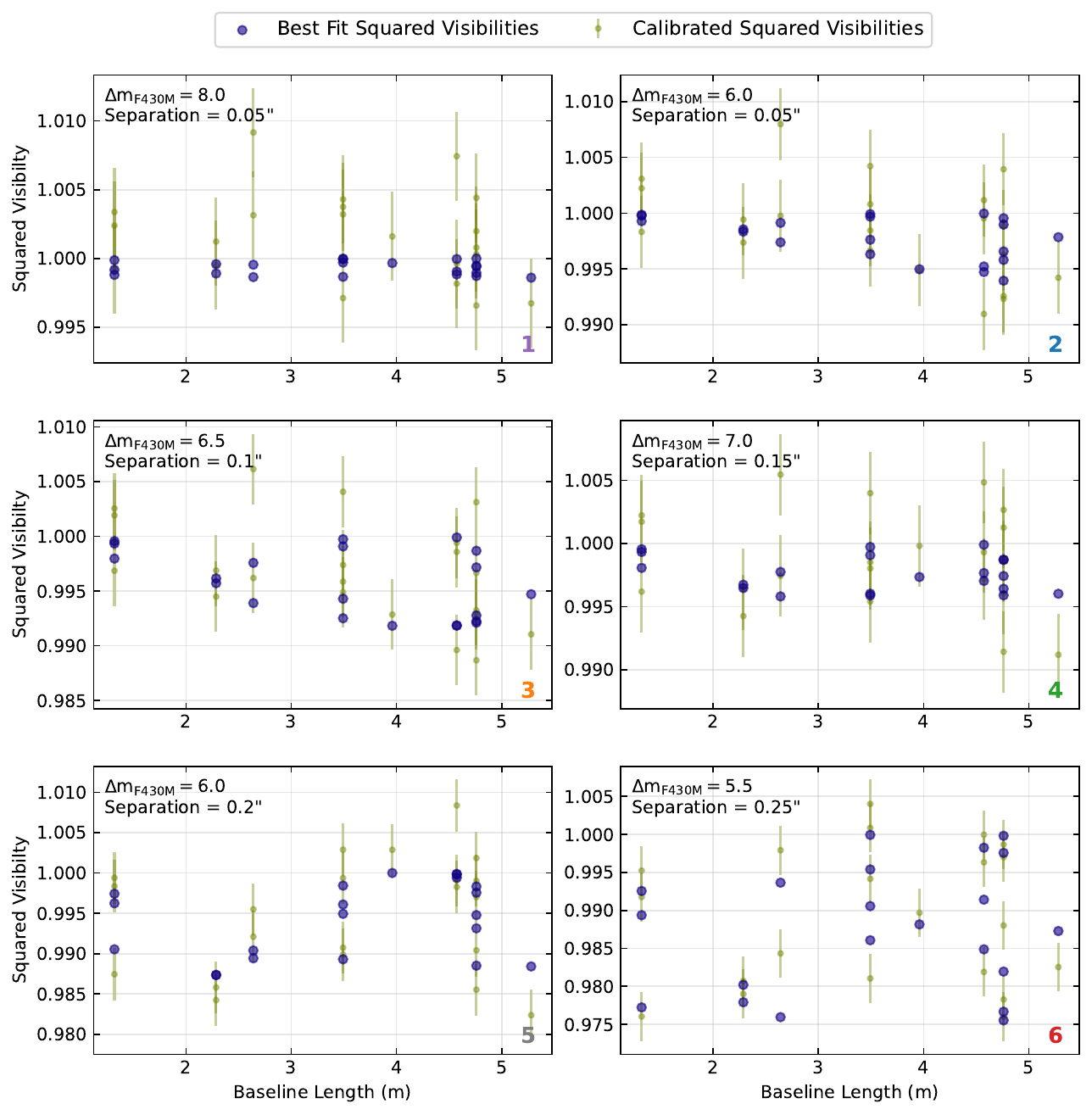}
    \caption{The calibrated squared visibilities and the corresponding best fits for the injected companion signals. The \textbf{\color{blue}test} numbers are colour coded according to Figure \ref{fig:ContrastsChisqComp} at the bottom right for each plot. The $\Delta m_{F430M}$ and separation values for each injection is also provided on the top left for each plot.}
    \label{fig:InjectedV2s}
\end{figure*}

\end{document}